\documentclass[twocolumn]{aastex631} 

\usepackage{amsmath}
\usepackage{empheq}
\usepackage{mathrsfs}
\usepackage{textcomp}
\usepackage{enumitem}   
\usepackage{gensymb}
\usepackage{hyperref}
\usepackage{graphicx}
\usepackage[caption=false]{subfig}
\usepackage{multirow}
\usepackage[flushleft]{threeparttable}
\usepackage{booktabs} 
\usepackage{color, colortbl} 

\hypersetup{colorlinks, linkcolor={blue}, citecolor={blue}, urlcolor={blue}} 

\definecolor{DarkOrange}{RGB}{204, 85, 0}
\definecolor{LincolnGreen}{RGB}{17, 102, 0}
\def\ion#1#2{#1$\;${\footnotesize\rm{#2}}\relax}

\newcommand{\ad}[1]{{ {#1}}}

\usepackage{xspace}

\newcommand\chandra{\textit{Chandra}\xspace}
\newcommand\swift{\textit{Swift}\xspace}
\newcommand\gaia{\textit{Gaia}\xspace}
\newcommand\xmm{\textit{XMM-Newton}\xspace}
\newcommand\maxi{\textit{MAXI}\xspace}
\newcommand\srg{\textit{SRG}\xspace}

\newcommand\galex{\textit{GALEX}\xspace}
\newcommand\target{AT2020mrf\xspace}

\def \caltech {{Division of Physics, Mathematics and Astronomy, 
California Institute of Technology, Pasadena, CA 91125, USA}}

\def \berkeley {{Department of Astronomy, University of California, Berkeley, 501 Campbell Hall, Berkeley, CA, 94720, USA}}

\def \iki {{Space Research Institute, Russian Academy of Sciences, Profsoyuznaya ul. 84/32, Moscow, 117997, Russia}}

\def \mpa {{Max-Planck-Institut f\"{u}r Astrophysik, Karl-Schwarzschild-Str. 1, D-85741 Garching, Germany}}

\shorttitle{AT2020mrf}
\shortauthors{Yao et al.}

\begin{document}
\pagenumbering{arabic}

\title{The X-ray and Radio Loud Fast Blue Optical Transient AT2020mrf: \\
Implications for an Emerging Class of Engine-Driven Massive Star Explosions}

\correspondingauthor{Yuhan Yao}
\email{yyao@astro.caltech.edu}

\author[0000-0001-6747-8509]{Yuhan Yao}
\affiliation{\caltech}

\author[0000-0002-9017-3567]{Anna Y. Q.~Ho}
\affiliation{\berkeley}
\affiliation{Miller Institute for Basic Research in Science, 468 Donner Lab, Berkeley, CA 94720, USA}

\author[0000-0002-9380-8708]{Pavel Medvedev}
\affiliation{\iki}

\author[0000-0002-8070-5400]{Nayana A. J.}
\affiliation{Indian Institute of Astrophysics, II Block, Koramangala, Bangalore 560034, India}

\author[0000-0001-8472-1996]{Daniel A.~Perley}
\affiliation{Astrophysics Research Institute, Liverpool John Moores University, IC2, Liverpool Science Park, 146 Brownlow Hill, Liverpool L3 5RF, UK}

\author[0000-0001-5390-8563]{S. R. Kulkarni}
\affiliation{\caltech}

\author[0000-0002-0844-6563]{Poonam Chandra}
\affiliation{National Centre for Radio Astrophysics, Tata Institute of Fundamental Research, P.O. Box 3, Pune, 411007, India}

\author{Sergey Sazonov}  
\affiliation{\iki}
\affiliation{Moscow Institute of Physics and Technology, Institutsky per. 9, 141700 Dolgoprudny, Russia}

\author{Marat Gilfanov}
\affiliation{\iki}
\affiliation{\mpa}

\author{Georgii Khorunzhev}
\affiliation{\iki}

\author[0000-0003-4307-0589]{David K. Khatami}
\affiliation{\berkeley}

\author{Rashid Sunyaev}
\affiliation{\iki}
\affiliation{\mpa}

\begin{abstract}
We present AT2020mrf (SRGe\,J154754.2$+$443907), an extra-galactic ($z=0.1353$) \ad{fast blue optical transient (FBOT) with a rise time of} $t_{g,\rm rise}=3.7$\,days \ad{and a peak luminosity of} $M_{g,\rm peak}=-20.0$. Its optical spectrum around peak shows a broad ($v\sim0.1c$) emission feature on a blue continuum ($T\sim2\times10^4$\,K), which bears a striking resemblance to AT2018cow. Its bright radio emission ($\nu L_\nu = 1.2\times 10^{39}\,{\rm erg\,s^{-1}}$; $\nu_{\rm rest}= 7.4$\,GHz; 261\,days) is similar to four other AT2018cow-like events, and can be explained by synchrotron radiation from the interaction between a sub-relativistic ($\gtrsim0.07$--$0.08c$) forward shock and a dense environment ($\dot M \lesssim 10^{-3}\,M_\odot \,{\rm yr^{-1}}$ for $v_{\rm w}=10^3\,{\rm km\,s^{-1}}$). AT2020mrf occurs in a galaxy with $M_\ast \sim 10^8\,M_\odot$ and specific star formation rate $\sim 10^{-10}\, {\rm yr^{-1}}$, supporting the idea that AT2018cow-like events are preferentially hosted by dwarf galaxies. The X-ray \ad{luminosity of AT2020mrf is the highest among FBOTs}. At 35--37\,days, \srg/eROSITA detected luminous ($L_{\rm X}\sim 2\times 10^{43}\,{\rm erg\,s^{-1}}$; 0.3--10\,keV) X-ray emission. The X-ray spectral shape ($f_\nu \propto \nu^{-0.8}$) and erratic intraday variability are reminiscent of AT2018cow, but the luminosity is a factor of $\sim20$ greater than AT2018cow. At 328\,days, \chandra detected it at $L_{\rm X}\sim10^{42}\,{\rm erg\,s^{-1}}$, which is $>200$ times more luminous than AT2018cow and CSS161010. At the same time, the X-ray emission remains variable on the timescale of $\sim1$\,day. We show that a central engine, probably a millisecond magnetar or an accreting black hole, is required to power the explosion. We predict the rates at which events like AT2018cow and AT2020mrf will be detected by \srg and Einstein Probe.
\end{abstract}

\vspace{1em}

\section{Introduction}
The past several years have shown that the landscape of massive-star death is unexpectedly rich and diverse. Of particular interest is the group of ``fast blue optical transients'' (FBOTs; \citealt{Drout2014, Pursiainen2018}). As implied by the name, these events exhibit blue colors of $(g-r)<-0.2$\,mag at peak, and evolve faster than ordinary supernovae (SNe), with time above half-maximum $t_{1/2}\lesssim 12$\,days. 

The earliest studies were stymied by the identification of FBOTs after the transients had faded away. This situation has been rectified by cadenced wide-field optical sky surveys such as the Zwicky Transient Facility (ZTF; \citealt{Bellm2019b}; \citealt{Graham2019}) and the Asteroid Terrestrial-impact Last Alert System (ATLAS; \citealt{Tonry2018}), which enable real-time discovery and spectroscopic classification. \citet{Ho2021a} recently identified three distinct subtypes of FBOTs: (1) subluminous stripped-envelope SNe of type Ib/IIb, (2) luminous interaction-powered SNe of type IIn/Ibn/Icn, and (3) the most luminous ($-20 \lesssim M_{\rm peak}\lesssim -22$) and short-duration ($t_{1/2}\lesssim 5$\,days) events with properties similar to AT2018cow. 

The nature of AT2018cow-like events remains the most mysterious. Following the discovery of the prototype AT2018cow ($z=0.014$, \citealt{Prentice2018}), only three analogs have been identified: AT2018lug ($z = 0.271$, \citealt{Ho2020}), CSS161010 ($z = 0.034$; \citealt{Coppejans2020}), and AT2020xnd ($z = 0.243$, \citealt{Perley2021xnd}). All of these events arose in low-mass star-forming galaxies, which suggests a massive star origin and disfavors models invoking tidal disruption by an intermediate-mass black hole \citep{Perley2021xnd}. In the radio and millimeter band, their high luminosities imply the existence of dense circumstellar material (CSM), which points to significant mass-loss prior to the explosion \citep{Ho2019, Huang2019, Margutti2019, Coppejans2020}. 

The X-ray luminosity of AT2018cow ($\sim10^{43}\,{\rm erg\,s^{-1}}$) at early times ($\lesssim 20$\,days) is similar to that of long-duration gamma-ray bursts (GRBs) \citep{RiveraSandoval2018}. Its fast soft X-ray variability suggests the existence of a central energy source (also called central engine), and the relativistic reflection features seen in the hard X-ray spectrum points to equatorial materials \citep{Margutti2019}. Probable natures of the central engine include an accreting black hole, a rapidly spinning magnetar, and an embedded internal shock \citep{Margutti2019, Pasham2021}. 
Meawhile, AT2018cow's late-time ($\sim20$--45\,days) optical spectra are dominated by hydrogen and helium \citep{Perley2019, Margutti2019, Xiang2021}, making it different from other engine-powered massive stellar transients such as long GRBs and hydrogen-poor super-luminous supernovae (i.e., SLSNe-I; see a recent review by \citealt{Gal-Yam2019}), which are devoid of hydrogen and helium. 

X-ray observations of AT2020xnd showed a luminosity consistent with that of AT2018cow at 20--40\,days \citep{Bright2022, Ho2021b}. Separately, late-time ($\gtrsim100$\,day) observations of AT2018cow and CSS161010 showed modest X-ray emission at $L_{\rm X}\approx {\rm few}\times 10^{39}\,{\rm erg\,s^{-1}}$) (see Appendix~\ref{sec:18cow_xmm} and \citealt{Coppejans2020}). AT2018cow-like events are thus promising X-ray transients to be discovered by the eROSITA \citep{Predehl2021} and the Mikhail Pavlinsky ART-XC \citep{Pavlinsky2021} telescopes on board the \textit{Spektrum-Roentgen-Gamma} (\srg) satellite \citep{Sunyaev2021}.

\target is an FBOT first detected by ZTF on 2020 June 12. On June 14, it was also detected by ATLAS. On July 15, it was reported to the transient name server (TNS) by the ATLAS team \citep{Tonry2020}. On June 17, an optical spectrum obtained by the ``Global SN Project'' displayed a featureless blue continuum. \citet{Burke2020} assigned a spectral type of ``SN II'', and tentatively associated it with a $z= 0.059$ host galaxy ($109^{\prime\prime}$ offset). \target was detected in the X-ray by \srg from 2020 July 21 to July 24 (\S\ref{subsec:srg_obs}), which made it a promising candidate AT2018cow analog and motivated our follow up observations. Given that the \srg detection occurred $\sim 41$ days after the first optical detection, and we became aware of it even later, our follow-up started in April 2021.

The paper is organized as follows. We outline optical, X-ray, and radio observations, as well as analysis of \target and its host galaxy ($z=0.1353$) in \S\ref{sec:obs}. 
We provide the forward shock and CSM properties in \S\ref{subsec:shock_CSM},
discuss possible power sources of the optical emission in \S\ref{subsec:optlc_model},
and present host galaxy properties in \S\ref{subsec:dwarf_host}.
We summarize \target's key X-ray properties and discuss the implication in the context of engine driven explosions similar to AT2018cow in \S\ref{subsec:xray_origin}.
We estimate the detection rates of events like AT2018cow and AT2020mrf in current and upcoming X-ray all-sky surveys in \S\ref{sec:rate}.
We give a summary in \S\ref{sec:conclusion}. 

UT time is used throughout the paper. We assume a cosmology of $\Omega_{\rm M} = 0.3$, $\Omega_{\Lambda}=0.7$, and $h=0.7$, implying a luminosity distance to \target of $D_L = 637\,{\rm Mpc}$ and an angular-diameter distance of $D_\theta = 494\,{\rm Mpc}$. Optical magnitudes are reported in the AB system. We use the Galactic extinction from \citet{Schlafly2011} and the extinction law from \citet{Cardelli1989}. Coordinates are given in J2000. 

\section{Observations and Data Analysis} \label{sec:obs}

\subsection{Optical Photometry} \label{subsec:opt_phot}

\begin{figure*}[htbp]
    \centering
    \includegraphics[width=0.7\textwidth]{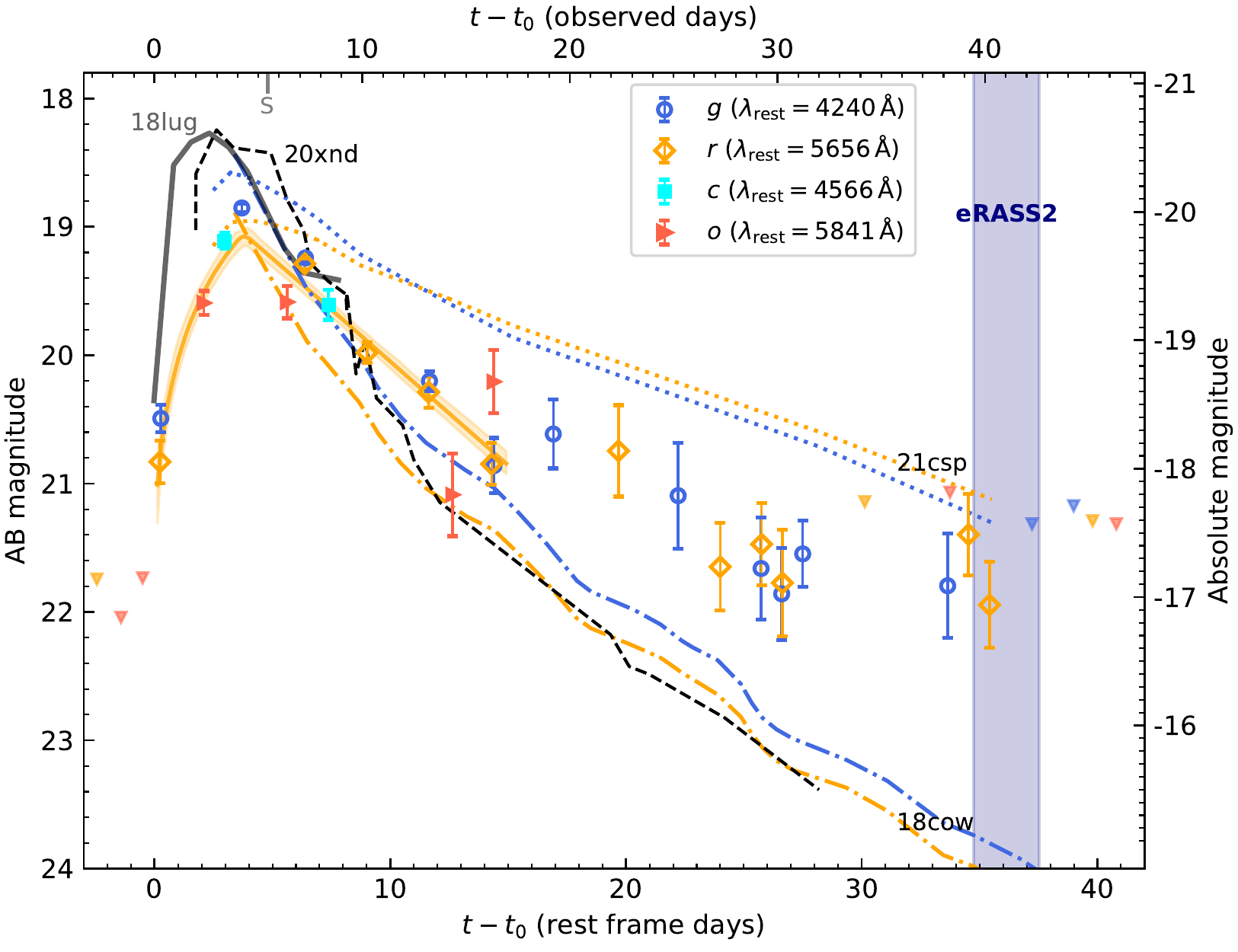}
    \caption{Optical (ZTF $gr$, ATLAS $co$) light curve of \target (data points are $>2.5\sigma$ detections, semi-transparent downward triangles are $3\sigma$ upper limits). The `S' tick along the upper axis marks the epoch of spectroscopy (\S\ref{subsec:opt_spec}). The solid orange line is a simple $r_{\rm ZTF}$-band model fitted to data round maximum (see \S\ref{subsec:optlc_model}). 
    The \srg eRASS2 scan duration is marked by the vertical blue band (\S\ref{subsec:srg_obs}). 
    The rest-frame equivalent light curves of AT2018cow and AT2021csp are shown as dash-dotted lines (based on blackbody parameters provided in Tab.~4 of \citealt{Perley2019}) and dotted lines (based on Tab.~4 of \citealt{Perley2022}), respectively. 
    The solid and dashed black lines are observer-frame $r_{\rm ZTF}$-band light curves of AT2018lug ($\lambda_{\rm rest}=5050$\,\AA) and AT2020xnd ($\lambda_{\rm rest}=5165$\,\AA), respectively. 
    Note that the apparent AB magnitude scale pertains to AT2020mrf only --- light curves of other objects are only shown in absolute magnitude.
    \label{fig:optlc}}
\end{figure*}

\begin{figure*}[htbp!]
    \centering
    \includegraphics[width=\textwidth]{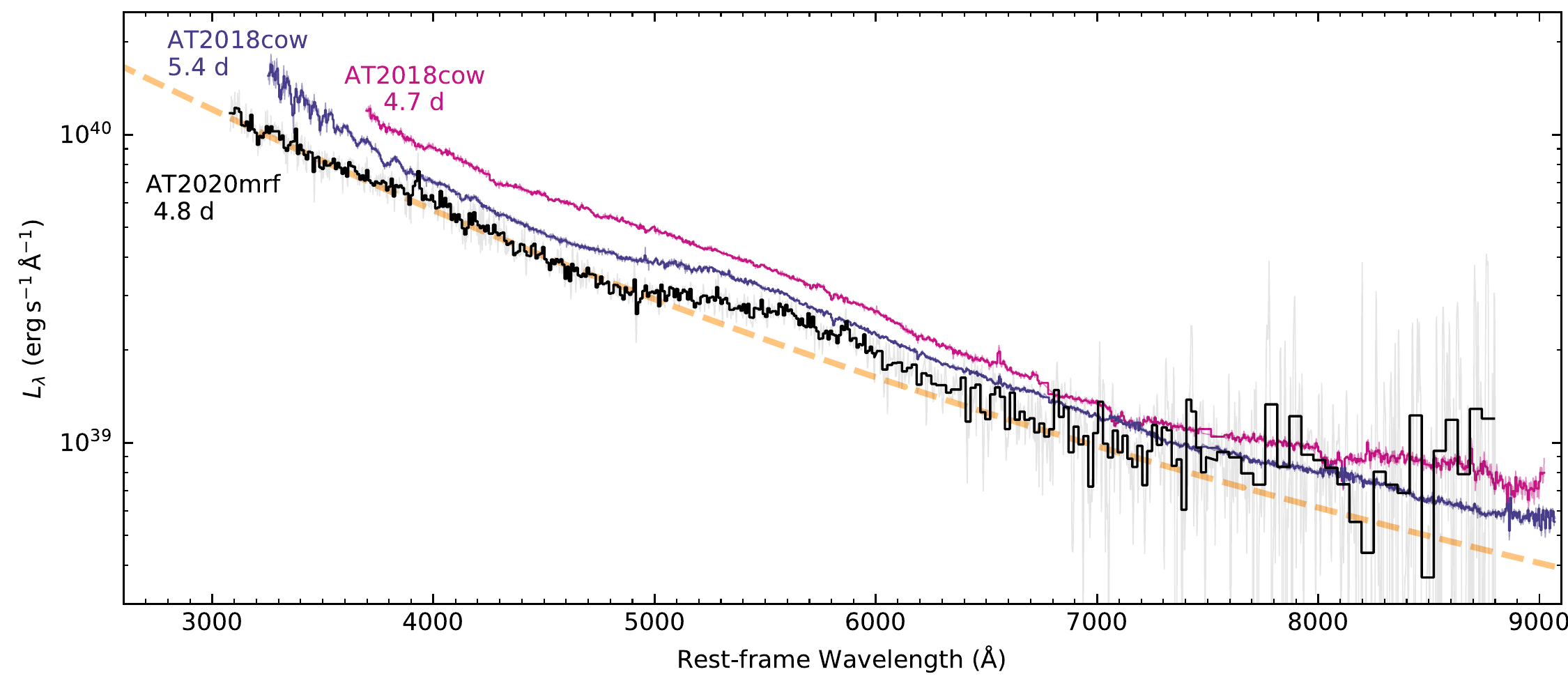}
    \caption{Optical spectrum of \target, compared with AT2018cow at similar phases \citep{Perley2019}. The dashed line is a blackbody with $T = 2\times10^4$\,K and $R = 7.9\times 10^{14}$\,cm. \label{fig:mrf_spec}}
\end{figure*}

We obtained public ZTF\footnote{\url{https://ztfweb.ipac.caltech.edu/cgi-bin/requestForcedPhotometry.cgi}} and ATLAS\footnote{\url{https://fallingstar-data.com/forcedphot/}} forced photometry \citep{Masci2019, Smith2020} using the median position of all ZTF alerts 
(${\rm R.A.}=15^{\rm h}47^{\rm m}54.17^{\rm s}$, 
${\rm decl.}=+44^{\circ}39^{\prime}07.34^{\prime\prime}$). The 1-day binned optical light curve is shown in Figure~\ref{fig:optlc}. Following \cite{Whitesides2017} and \citet{Ho2020}, we compute absolute magnitude using 
\begin{align}
    M = m_{\rm obs} - 5\,{\rm log_{10}}\left( \frac{D_L}{\rm 10\,pc} \right) + 2.5 \,{\rm log_{10}} (1+z) \label{eq:Mabs}
\end{align}
The last term in Equation~(\ref{eq:Mabs}) is a rough estimation of $K$-correction, and introduces an error of 0.1\,mag.

The first detection is $r = 20.88\pm 0.17$, on 2020-06-12T06:14:12 (59012.2599\,MJD) and the last non-detection is $o > 21.73$, on 2020-06-11T10:12:13 (59011.4252\,MJD). Therefore, we assume an explosion epoch of $t_0 = 59012.0$\,MJD. Hereafter we use $\Delta t$ to denote rest-frame time with respect to $t_0$. At $\Delta t = 3.7$\,days, \target peaked at $M_{g}=-20.0$\,mag.

\subsection{Optical Spectroscopy}\label{subsec:opt_spec}

The transient spectrum\footnote{Available at \url{https://www.wis-tns.org/object/2020mrf}} was obtained on 2020 June 17 ($\Delta t = 4.8$\,days) with the FLOYDS-N spectrograph on the 2\,m Faulkes Telescope North \citep{Burke2020}. As shown in Figure~\ref{fig:mrf_spec}, the spectrum is similar to that of AT2018cow at similar phases --- a single broad feature at $\sim 5600$\,\AA\ was observed to span $\pm 600$\,\AA, indicating a velocity of $0.1c$. 
\ad{The origin of this broad line in AT2018cow remains an open question.
\citet{Perley2019} note that although it is vaguely reminiscent of the \ion{Fe}{II} feature in Ic broad-line (Ic-BL) SNe around peak \citep{Galama1998}, in SNe Ic-BL the blueshifted absorption trough strengthens at later times, while in AT2018cow this line vanished at $\Delta t\sim8$\,days.
In terms of other AT2018cow-like objects, the peak-light optical spectra of AT2018lug and AT2020xnd are consistent with being blue and featureless \citep{Ho2020, Perley2021xnd}, and that there exists no published optical spectra of CSS161010.}

A blackbody fit to \target's optical spectrum suggests a temperature of $T \approx 2\times 10^{4}$\,K and a radius of $R = 7.9\times 10^{14}$\,cm. This temperature is typical of AT2018cow-like events.

\subsection{Early-time X-rays: \srg} \label{subsec:srg_obs}

\srg is a space satellite at the L2 Lagrange point with a drafting rate of $\approx 1^{\circ}\,{\rm \,day^{-1}}$. It is conducting eight all-sky surveys from the beginning of 2020 to the end of 2023, with a cadence of 6\,months. Hereafter eRASS$n$ refers to the $n$'th eROSITA all-sky survey. \srg's rotational axis points toward the Sun, and the rotational period is 4\,hours. The eROSITA field-of-view (FoV) is 1\,deg$^2$. Therefore, during a single sky survey, a particular region of the sky will be scanned by eROSITA at least $\sim6$ times ($\sim1$\,day), where each scan lasts for $\approx40$\,s (see details in \citealt{Sunyaev2021}). \target, at a relatively high ecliptic latitude of $b_{\rm ecl} = 61.9^{\circ}$, was scanned for $\approx 3$\,days in each all-sky survey. 

\begin{figure}
    \centering
    \includegraphics[width=\columnwidth]{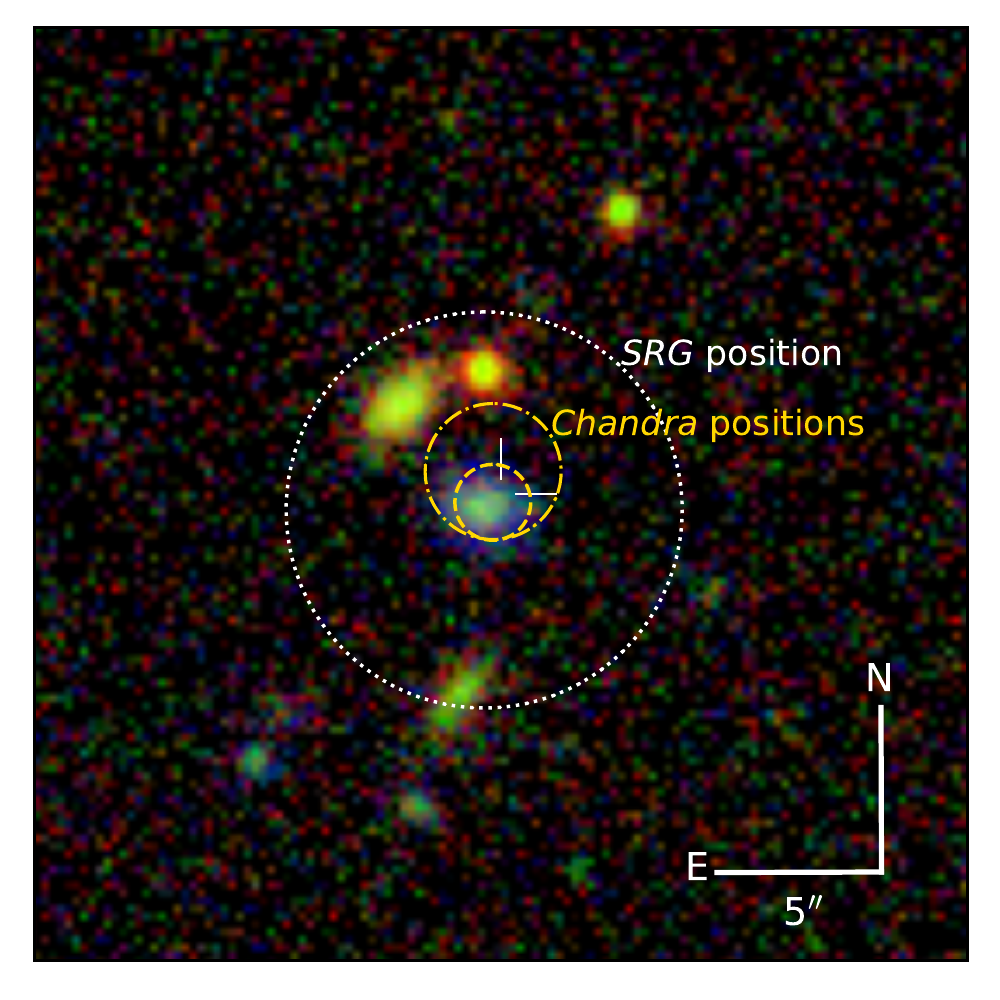}
    \caption{HSC-SSP RGB false-color $g$/$i$/$z$ image centered at the ZTF position of \target (marked by the white crosshairs). \target is an off-nuclear source 0.50$^{\prime\prime}$ offset from the host centroid. \ad{The position of the X-ray transient detected by \srg/eROSITA is shown with a dotted circle, where the radius represents the $4.28^{\prime\prime}$ uncertainty (98\% confidence). The more accurate positions provided by \chandra obsID 25050 and obsID 25064 are shown with a dash-dotted circle and a dashed circle (1$\sigma$ confidence), respectively. }
    \label{fig:host_image}}
\end{figure}

\begin{figure}[htbp!]
    \centering
    \includegraphics[width=\columnwidth]{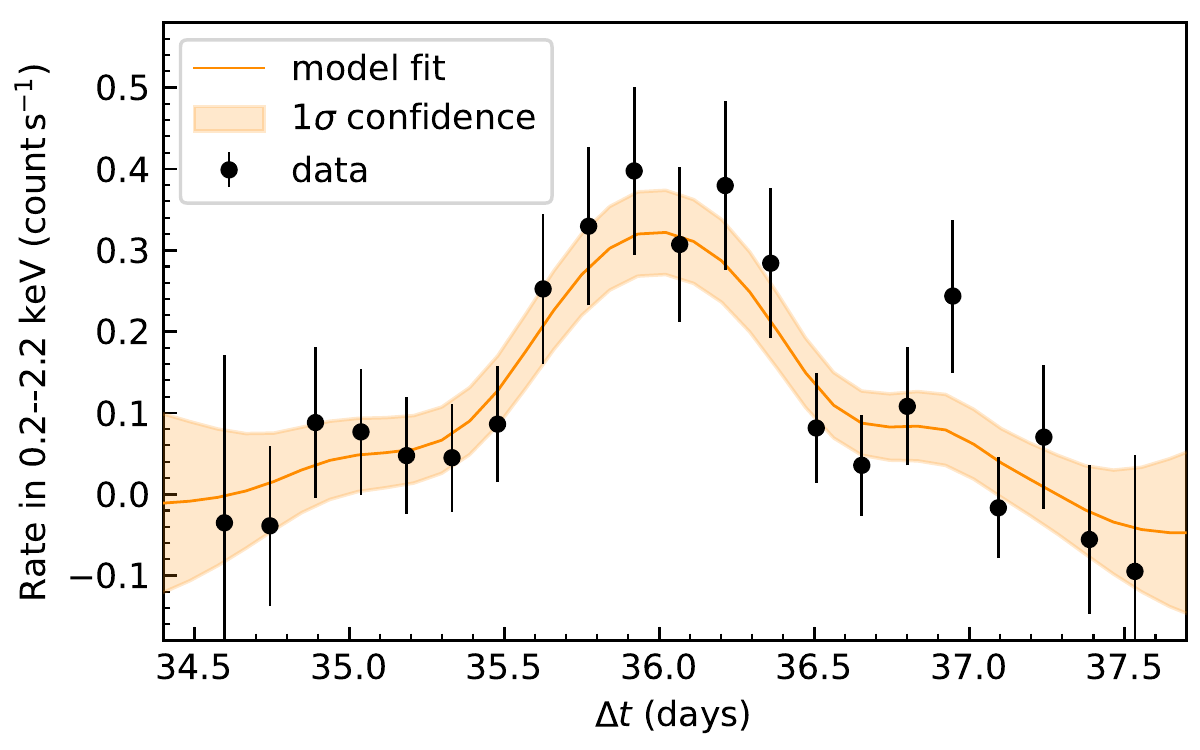}
   \caption{eRASS2 light curve of \target. Count rate uncertainties are estimated using Gehrels' approximation \citep{Gehrels1986}. 
   The orange curve is a fit to the data, generated using a Gaussian process model following procedures laid out in Appendix B.4 of \citet{Yao2020}. 
   \label{fig:srg_xlc}}
\end{figure}

During eRASS2 ($\Delta t \sim 36$\,days), \srg/eROSITA discovered an X-ray transient SRGe\,J154754.2$+$443907, with a 98\% localization radius of $4.28^{\prime\prime}$. 
SRGe\,J154754.2$+$443907 is only 0.56$^{\prime\prime}$ from \target\ \ad{(see Figure~\ref{fig:host_image})}, suggesting an association between the X-ray and the optical transients. 
Figure~\ref{fig:srg_xlc} shows that the source exhibits significant variability --- the 0.2--2.2\,keV count rate increased from $\approx 0.053\,{\rm count\,s^{-1}}$ ($\Delta t\sim 35$\,days) to $\approx 0.32\,{\rm count\,s^{-1}}$ ($\Delta t\sim 36$\,days), and then decreased to $\approx 0.051\,{\rm count\,s^{-1}}$ ($\Delta t\sim 37$\,days). 

\begin{figure}[htbp!]
    \centering
    \includegraphics[width=\columnwidth]{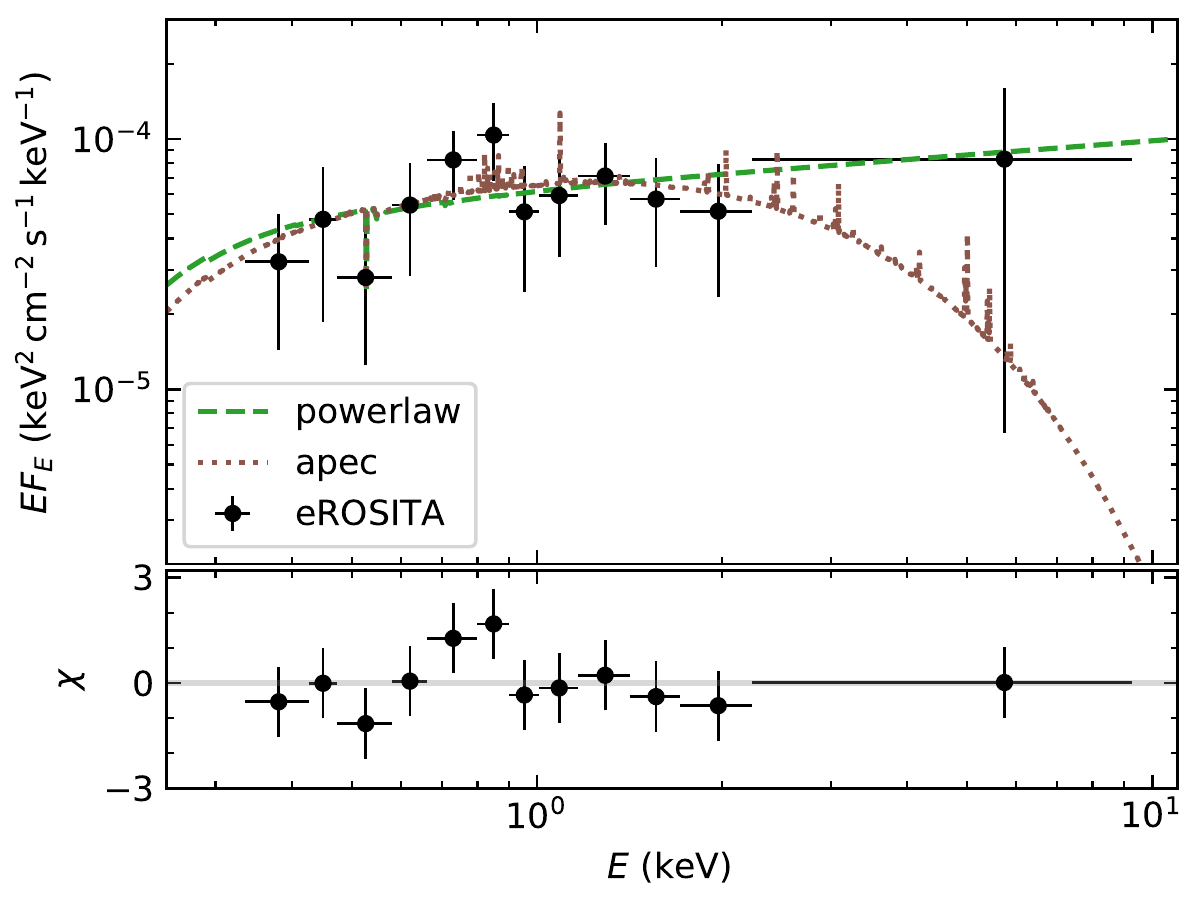}
    \caption{eRASS2 spectrum of \target. We show the best-fit power-law model (dashed green line, $f_\nu \propto \nu^{-0.8}$) and thermal plasma model (dotted brown line, $k_{\rm B}T\sim 2.0$\,keV) with $N_{\rm H}$ fixed at the Galactic value (see Table~\ref{tab:xfit}). \label{fig:srg_spec}}
\end{figure}

\begin{deluxetable*}{@{\extracolsep{4pt}}cccccc}[htbp!]
    \tablecaption{Modeling of the eRASS2 spectrum.\label{tab:xfit}}
	\tablehead{
    \colhead{Component}
	& \colhead{Parameter}
	& \multicolumn{2}{c}{Power-law Model}
	& \multicolumn{2}{c}{Thermal Plasma Model}\\
	\cline{3-4}
	\cline{5-6}
	\colhead{}
	& \colhead{}
	& \colhead{(a) Fixed $N_{\rm H}$}
	& \colhead{(b) Free $N_{\rm H}$}
	& \colhead{(a) Fixed $N_{\rm H}$}
	& \colhead{(b) Free $N_{\rm H}$}
	}
	\startdata
	\texttt{tbabs} &$N_{\rm H}$ ($10^{20}\,{\rm cm^{-2}}$) 
	& 1.38 & $15.32^{+14.06}_{-10.70}$
	& 1.38 & $9.04^{+7.10}_{-6.99}$\\
	\texttt{zpowerlw} & $\Gamma$ 
	& $1.81\pm0.26$ 
	& $2.79^{+1.00}_{-0.80}$ & ... & ... \\
	& $\rm norm_{pl}$ ($10^{-5}$) 
	& $8.0_{-1.0}^{+1.1}$ 
	& $14.8^{+11.8}_{-5.7}$ &... &... \\
	\texttt{apec} & $k_{\rm B}T$ & ... & ... 
	& $2.0_{-0.7}^{+1.9}$ 
	& $1.0^{+1.1}_{-0.3}$ \\
	              & $\rm norm_{apec}$ ($10^{-4}$) & ... & ... 
	              & $3.8_{-0.7}^{+0.9}$ 
	              & $7.2_{-3.3}^{+5.2}$ \\
	\hline
	\multicolumn{2}{c}{\textit{cstat}/\textit{dof}} & 25.94/35 & 24.09/34 & 24.80/35 & 23.60/34 \\
	\multicolumn{2}{c}{Observed 0.3--10\,keV flux ($10^{-13}\,{\rm erg\,s^{-1}\,cm^{-2}}$)}  
	& $3.90_{-1.00}^{+1.32}$ 
	& $2.48_{-0.83}^{+0.51}$
	& $2.40_{-0.75}^{+0.54}$	
    & $1.92_{-1.37}^{+0.15}$
	\enddata
\tablecomments{
$\rm norm_{pl}$ and $\rm norm_{apec}$ are the normalization parameters in the model components (see the \texttt{xspec} documentation for units). Uncertainties are represented by the 68\% confidence intervals. }
\end{deluxetable*}

Figure~\ref{fig:srg_spec} shows the average eRASS2 spectrum of \target, which has been grouped via \texttt{ftgrouppha} to have at least five counts per bin in the background spectrum. 
We fit the 0.3--10\,keV spectrum using \texttt{xspec} (12.11, \citealt{Arnaud1996}) and $C$-statistics. 
The data are modeled first with an absorbed power-law (\texttt{zpowerlw}) and then with an absorbed thermal plasma (\texttt{apec}). For each model, we first fix the column density at the Galactic value of $N_{\rm H} = 1.38\times 10^{20}\,{\rm cm^{-2}}$ \citep{Willingale2013}, and then free this parameter. The models with fixed $N_{\rm H}$ are shown in Table~\ref{tab:xfit}. The data do not favor any particular model, since the \textit{cstat}/\textit{dof} ($C$-statistics divided by degrees of freedom) values have small differences between the four fits.

Although we are not able to distinguish between the power-law and thermal models using the eROSITA data, the optical/radio similarities between AT2020mrf and AT2018cow (\S\ref{subsec:opt_phot}, \S\ref{subsec:opt_spec}, \S\ref{subsec:shock_CSM}), and the non-thermal nature of AT2018cow's X-rays ($f_\nu \propto \nu^{-0.7}$, 36.5\,days, 0.3--30\,keV, see Fig.~6 of \citealt{Margutti2019}) motivate us to adopt the power-law model in the following discussion.


\target was not detected in eRASS1, eRASS3 and eRASS4. 
Using the eROSITA sensitivity maps, we calculate the 0.3--2.2\,keV flux upper limits to be $(1.12, 1.35, 1.54)\times 10^{-14}\,{\rm erg\,s^{-1}\,cm^{-2}}$ at the confidence level likelihood of 6 ($\approx 2.8\sigma$).

\subsection{Late-time X-rays: \chandra} \label{subsec:cxo}

We conducted deep X-ray observations of \target with the \chandra X-ray Observatory \citep{Chandra2020} under a DDT program (PI Yao) on 2021 June 18 (22.0\,ks, obsID 25050) and June 19 (19.8\,ks, obsID 25064). We used the Advanced CCD Imaging Spectrometer (ACIS; \citealt{Garmire2003}), with the aim point on the back illuminated CCD S3. The data were reduced with the \texttt{CIAO} package (v4.14). 

\ad{In order to determine the astrometric shifts of \chandra images, we first ran the \texttt{CIAO} tool \texttt{wavdetect} to obtain lists of positions for all sources in the \chandra FoV. Wavelet scales of 1, 2, 4, and 8 pixels and a significance threshold of $10^{-6}$ were used. A total of 8 and 12 X-ray sources were detected in obsID 25050 and obsID 25064, respectively. We cross matched the X-ray source lists with the \gaia EDR3 catalog \citep{Gaia2021}, using a radius of $2^{\prime\prime}$. This left two \chandra/\gaia sources from both obsIDs. We define the astrometric shifts as the mean difference in R.A. and decl. between the two matched sources. 
For obsID 25050, 
$\delta {\rm R.A.}= -1.88\pm0.42^{\prime\prime}$ and 
$\delta {\rm decl.}= -0.58 \pm 0.75^{\prime\prime}$; 
For obsID 25064, 
$\delta {\rm R.A.}= -0.62\pm0.27^{\prime\prime}$ and 
$\delta {\rm decl}= +0.61 \pm 0.29^{\prime\prime}$.}

\ad{Having applied the astrometric shifts, we found that an X-ray source at the location of \target was detected in both obsIDs. The position of the X-ray source from obsID 25050 is 
${\rm R.A.}=15^{\rm h}47^{\rm m}54.18^{\rm s}$, 
${\rm decl.}=+44^{\circ}39^{\prime}07.83^{\prime\prime}$,
with an astrometric uncertainty of 1.47$^{\prime\prime}$ from the residual offsets with the \gaia catalog;
The position of the X-ray source from obsID 25064 is 
${\rm R.A.}=15^{\rm h}47^{\rm m}54.18^{\rm s}$, 
${\rm decl.}=+44^{\circ}39^{\prime}07.16^{\prime\prime}$,
with an astrometric uncertainty of 0.82$^{\prime\prime}$ from the residual offsets with the \gaia catalog.
The \chandra positions are shown in Figure~\ref{fig:host_image}, which are more accurate than the eROSITA position, and clearly associate the X-ray emission with the ZTF position of \target.
}

\ad{For each obsID, we extracted the source spectrum using a source region of $r_{\rm src} = 1.5^{\prime\prime}$ centered on the X-ray position determined by \texttt{wavdetect}. A total of 30 and 10 counts (0.5--10\,keV) were detected within the source regions of obsID 25050 and obsID 25064, respectively.
The background spectrum was extracted using nearby source-free regions. 
The 0.5--10\,keV net count rate at 90\% credible interval is $1.61^{+0.32}_{-0.28}\times10^{-3}\,{\rm count\,s^{-1}}$ for obsID 25050, and $0.56^{+0.21}_{-0.17}\times10^{-3}\,{\rm count\,s^{-1}}$ for obsID 25064, indicating that X-ray net count rate has dropped by a factor of $2.9\pm1.1$. Such a large flux decrease reflects intrinsic X-ray variability.}

\begin{figure}[htbp!]
    \centering
    \includegraphics[width=\columnwidth]{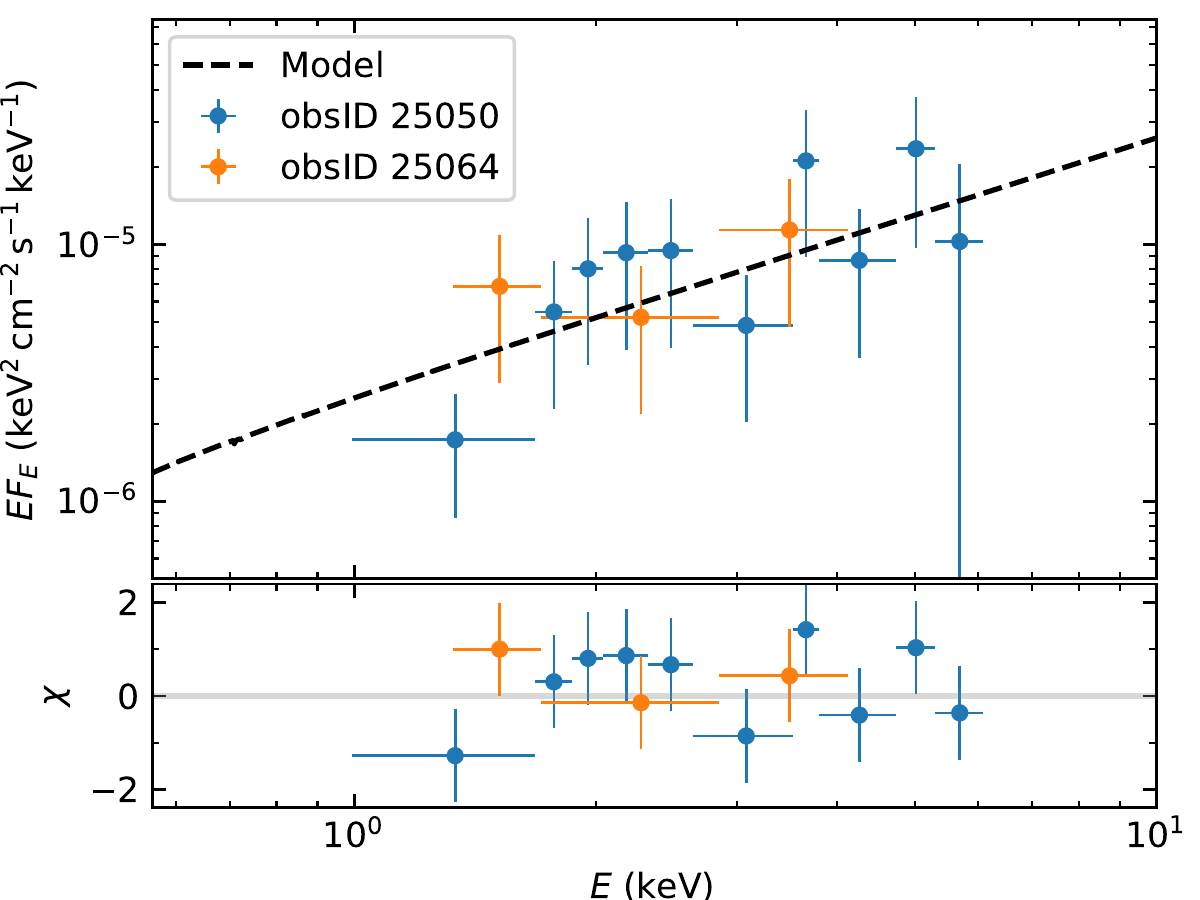}
    \caption{\chandra spectrum of \target at $\Delta t \approx 328$\,days. The data have been rebinned for visual clarity. The dashed line is the best-fit model ($f_\nu \propto \nu^{+0.0}$). To account for the flux variation (see text), the obsID 25064 data has been divided by 0.39. \label{fig:cxo_spec}}
\end{figure}

We groupped the \chandra spectrum to at least one count per bin, and \ad{modeled the 
data using $C$-statistics}. We used a model of \texttt{tbabs*zpowerlw}, with $N_{\rm H}$ fixed at the Galactic value. Since the count rate has significantly decreased between the two obsIDs, we include a constant scaling factor $\mathcal{C}$ between the two \chandra observations \citep{Madsen2017}, with the constant for obsID 25050 ($\mathcal{C}_1$) fixed at 1. \ad{The result, with $cstat/dof = 32.25/34$, gives $\Gamma = 1.00\pm0.35$ and $\mathcal{C}_2 = 0.39^{+0.17}_{-0.13}$, where uncertainties are represented by the 68\% confidence intervals. 
The best-fit model is shown in Figure~\ref{fig:cxo_spec}}. 

\begin{deluxetable}{cccc}[htbp!]
    \tablecaption{X-ray flux measurements of \target.\label{tab:xlc}}
	\tablehead{
	\colhead{$\Delta t$}
	& \colhead{Telescope}
	& \colhead{Observed 0.3--10\,keV flux} \\
	\colhead{(days)}
	& \colhead{}
	& \colhead{($10^{-14}\,{\rm erg\,s^{-1}\,cm^{-2}}$)}
	}
	\startdata
    $-127$ & \srg/eRASS1 & $<2.93$ \\
    34.5--37.6 & \srg/eRASS2 & $39.0_{-10.0}^{+13.2}$\\
    \hline
    192 & \srg/eRASS3 & $<7.24$\\
    \hline
    327.4 & \multirow{2}{*}{\chandra} & \ad{$4.00_{-1.24}^{+0.68}$}\\
    328.2 &  & \ad{$1.57_{-0.49}^{+0.27}$}\\
    \hline
    355 & \srg/eRASS4 & $<8.26$ \\
	\enddata
	\tablecomments{To convert the 0.3--2.2\,keV eROSITA upper limits to 0.3--10\,keV, we assume the eRASS2 best-fit spectral model for the eRASS1 epoch, and the \chandra spectral model for the eRASS3 and eRASS4 epochs. }
\end{deluxetable}

\begin{figure*}[htbp!]
    \centering
    \includegraphics[width=0.8\textwidth]{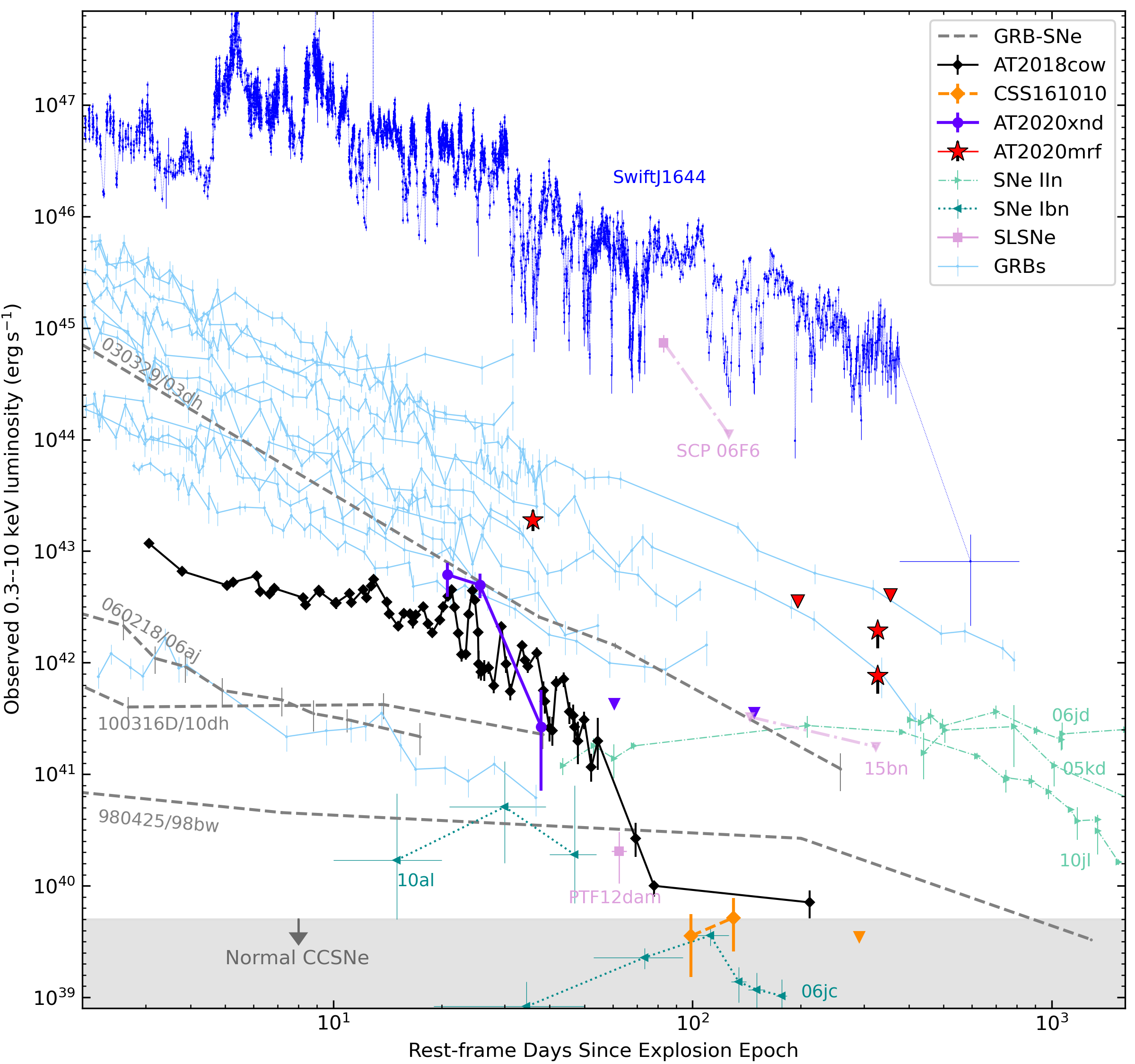}
\caption{X-ray emission of \target, compared with AT2018cow (\citealt{RiveraSandoval2018, Margutti2019}, Appendix~\ref{sec:18cow_xmm}), CSS161010 \citep{Coppejans2020}, AT2020xnd \citep{Bright2022, Ho2021b}, cosmological long GRBs (light blue solid lines; Appendix~\ref{sec:grb_xlc}), GRBs associated with SNe (dashed grey lines; \citealt{Kouveliotou2004, Tiengo2004, Campana2006, Soderberg2006, Margutti2013}), SLSNe-I (\citealt{Levan2013, Margutti2018}), the jetted TDE SwiftJ1644 \citep{Mangano2016}, interacting SNe of type IIn (dashed-dotted green lines; \citealt{Chandra2012, Chandra2015, Dwarkadas2016, Katsuda2016}) and type Ibn (dotted cyan lines; \citealt{Immler2008, Ofek2013}), as well as normal CCSNe \citep{Dwarkadas2012}. \target is as luminous as cosmological GRBs. 
    \label{fig:xlc}}
\end{figure*}

The difference between the \srg and \chandra power-law indices is $\Gamma_{\rm 36\,d} - \Gamma_{\rm 328\,d} = 0.81\pm0.44$. Therefore, we conclude that a change of $\Gamma$ is marginally detected at $1.9\sigma$. 
Table~\ref{tab:xlc} summarizes the 0.3--10\,keV fluxes. 

Figure~\ref{fig:xlc} compares the X-ray luminosity evolution of \target with other types of explosions. \ad{We further discuss this figure in \S\ref{subsec:xray_origin}.}

\subsection{Search for Prompt $\gamma$-rays} \label{subsec:gamma-ray}
Given that cosmological long GRBs are the only type of massive-star explosion with X-ray luminosities known to be  comparable to \target (see Figure~\ref{fig:xlc}), we are motivated to search for bursts of prompt $\gamma$-rays between the last ZTF non-detection and the first ZTF detection (\S\ref{subsec:opt_phot}). 
\ad{
During this time interval, only one burst was detected by the interplanetary network (IPN; \citealt{Hurley2010}). The position of this burst \citep{Sonbas2020} is inconsistent with that of AT2020mrf. To obtain a constraint on the $\gamma$-ray flux of \target, we use the Konus instrument \citep{Aptekar1995} on the \textit{Wind} spacecraft. 
Unlike other high energy telescopes on low Earth orbit (LEO) spacecrafts (such as \swift/BAT and \textit{Fermi}/GBM), Konus-\textit{Wind} (KW) continuously observe the whole sky without Earth blocking and with a very stable background, thanks to its orbit around the L1 Lagrange point (see, e.g., \citealt{Tsvetkova2021}). During the interval of interest, KW was taking data (total duration of data gaps was $<1$\% of the total time). Assuming a typical long GRB spectrum\footnote{\ad{The Band function with peak energy $E_{\rm peak}=300$\,keV, low-energy photon index $\alpha = -1$, and high energy photon index $\beta = -2$ \citep{Band1993}. }} and a timescale of 2.944\,s, KW gives a 20--1000\,keV upper limit of $<2\times10^{-7}\,{\rm erg\,s^{-1}\,cm^{-2}}$. This corresponds to an isotropic luminosity of $L_{\rm iso}<1.0\times 10^{49}\,{\rm erg\,s^{-1}}$, 
}
which strongly disfavors an on-axis classical GRB \citep{Frail2001}.

\subsection{Radio: VLA and uGMRT} \label{subsec:radio_obs}

\begin{deluxetable}{cc|crr}[htbp!]
	\tablecaption{Radio observations of \target.\label{tab:vla}}
	\tablehead{
		\colhead{Date}   
		& \colhead{$\Delta t$ }
		& \colhead{Telescope/} 
		& \colhead{$\nu_0$} 
		& \colhead{$f_\nu$}\\
		\colhead{in 2021}   
		& \colhead{(days)}
		& \colhead{Receiver} 
		& \colhead{(GHz)}
		& \colhead{($\mu$Jy)}
	}
	\startdata
\multirow{6}{*}{Apr 2} & \multirow{6}{*}{259.5}& \multirow{6}{*}{VLA/C} 
  & 4.30 & $254 \pm 25$ \\ 
&&& 4.94 & $ 234 \pm 22$ \\ 
&&& 5.51 & $ 330 \pm 18$ \\ 
&&& 6.49 & $ 327 \pm 20$ \\ 
&&& 7.06 & $ 336 \pm 17 $\\ 
&&& 7.70 & $ 349 \pm 18 $\\ 
\hline
\multirow{9}{*}{Apr 6} & \multirow{9}{*}{262.9}& VLA/S 
  & 3.00 & $165 \pm 26$ \\ 
 \cline{3-5}
 && \multirow{3}{*}{VLA/X} 
   & 8.49  & $277 \pm 23$ \\ 
 &&& 9.64  & $271 \pm 20$ \\ 
 &&& 11.13 & $223 \pm 17$ \\ 
 \cline{3-5}
 && \multirow{3}{*}{VLA/Ku} 
   & 12.78 & $213 \pm 19$ \\ 
 &&& 14.32 & $189 \pm 16$ \\ 
 &&& 16.62 & $153 \pm 15$ \\ 
 \cline{3-5}
 &&  \multirow{2}{*}{VLA/K}
   & 20.00  & $149 \pm 8$ \\ %
 &&& 24.00  & $103 \pm 8$ \\ %
\hline
\hline
May 19 & 300.9 & uGMRT/B5 &1.25 &  $<45$ \\
 \hline
May 29 & 309.5 & VLA/S & 3.00 & $206 \pm 48$ \\ 
\hline
\hline
Aug 13 & 376.6 & uGMRT/B5 &1.25 & $<105$ \\
\hline
\hline
Sep 28 & 416.8 & uGRMT/B5 & 1.36 & $68\pm15$ \\
\hline
\multirow{5}{*}{Sep 28--29}  & \multirow{5}{*}{417.5}  &VLA/S & 3.00 & $81\pm10$\\
        \cline{3-5}
        &   &VLA/C & 6.00 & $87\pm7$\\
        \cline{3-5}
        &  & VLA/X & 10.00  &$49\pm8$ \\
        \cline{3-5}
        &   & \multirow{2}{*}{VLA/Ku} & 13.55 & $65\pm6$ \\
        &   &       & 16.62 & $51\pm7$ \\
\enddata
\tablecomments{$\nu_0$ is observed central frequency. $f_\nu$ is the observed flux density values. Upper limits are $3\sigma$.}
\end{deluxetable}

We began a monitoring program of \target using the VLA \citep{Perley2011} under Program 21A-308 (PI Ho), and the upgraded Giant Metrewave Radio Telescope (uGMRT; \citealt{Swarup1991, Gupta2017}) under Program 40\_077 (PI Nayana). The data were analyzed following the standard radio continuum image analysis procedures in the Common Astronomy Software Applications (\texttt{CASA}; \citealt{McMullin2007}). The results are presented in Table~\ref{tab:vla}.
Incidentally, \target was not detected in the Karl G. Jansky Very Large Array Sky Survey (VLASS, \citealt{Lacy2020}), which provides a 3-$\sigma$ upper limit of 0.42\,mJy at 2--4\,GHz in March 2019. 
Hereafter radio flux density values have been $K$-corrected and frequency values are reported in the rest-frame.
\ad{$K$-correction was performed following \citet{Condon2018}, assuming a steep synchrotron spectrum with a spectral index of $\beta=-1$ ($f_\nu \propto \nu^{\beta}$). }

Regarding data obtained within $(\Delta t /10)$\,days as coeval, we model the radio spectral energy distribution (SED) at $\Delta t\approx 261$\,days and $\Delta t\approx 417 $\,days with a broken power-law \citep{Granot2002}:
\begin{align}
    L_\nu = L_{\nu\,\rm peak} \left[ \left( \frac{\nu}{\nu_{\rm peak}}\right)^{-s \beta_1} + \left( \frac{\nu}{\nu_{\rm peak}}\right)^{-s \beta_2} \right]^{-1/s} \label{eq:GSbrokenPL}
\end{align}
where $\nu$ and $L_\nu$ are quantities in the object's rest-frame, $L_{\nu\,\rm peak}$ is the peak specific luminosity, $\nu_{\rm peak}$ is the peak frequency, $\beta_1$ and $\beta_2$ are the asymptotic spectral indices below and above the break, and $s$ is a smoothing parameter. 
We perform the fit using the Markov chain Monte Carlo (MCMC) approach with \texttt{emcee} \citep{Foreman-Mackey2013}. The reported uncertainties follow from the 68\% credible region. 

\begin{figure}[htbp!]
    \centering
    \includegraphics[width = \columnwidth]{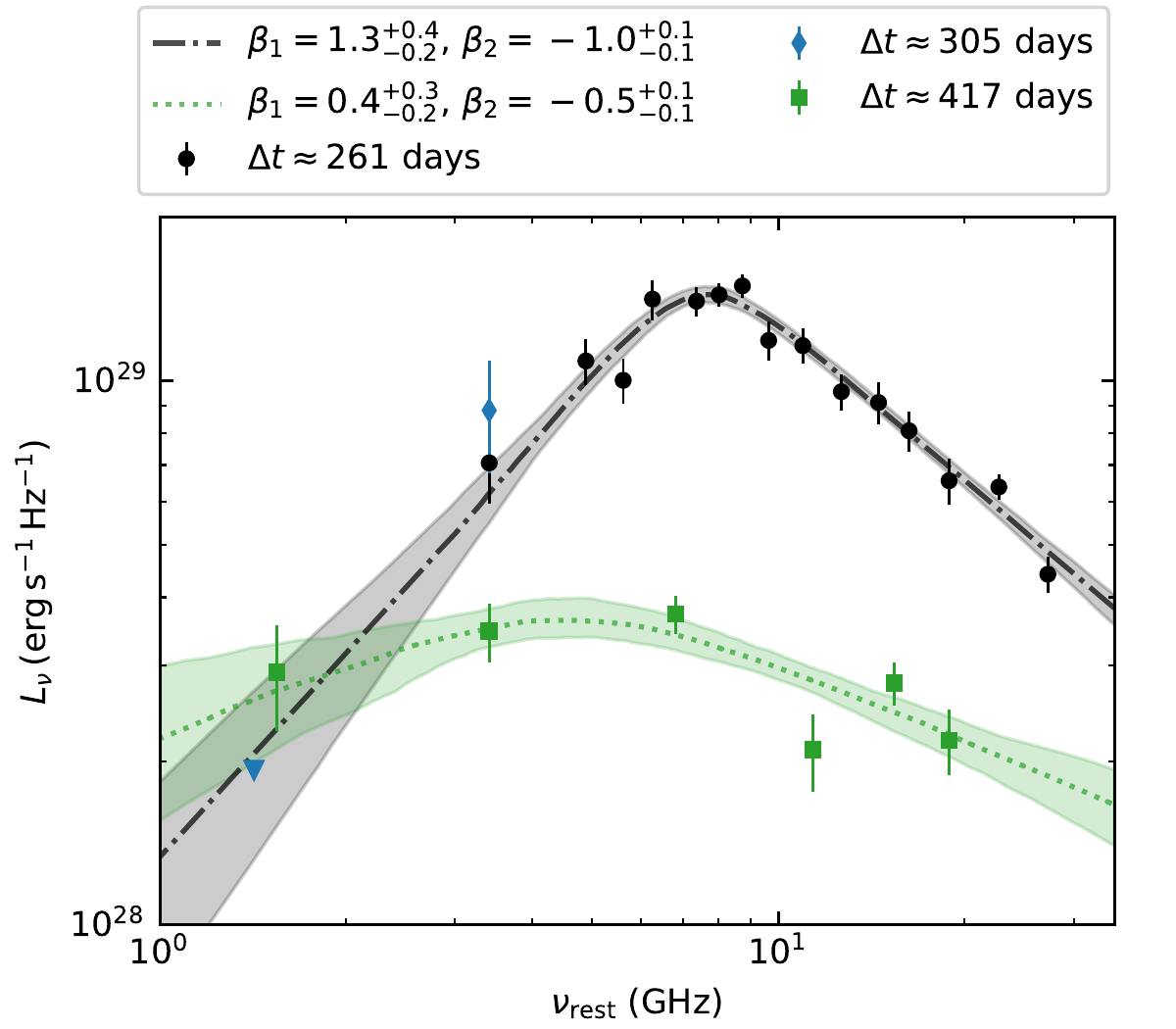}
    \caption{Radio observations of \target, overplotted with the best-fit broken power-law models. \label{fig:radio_sed}}
\end{figure}

The best-fit models are shown in Figure~\ref{fig:radio_sed}. 
At $\Delta t \approx 261$\,days, $\nu_{\rm peak} = 7.44_{-0.52}^{+0.44}$\,GHz, $L_{\nu\,\rm peak} = 1.70_{-0.09}^{+0.23}\times 10^{29}\,{\rm erg\,s^{-1}\,Hz^{-1}}$, $\beta_1 = 1.3^{+0.4}_{-0.2}$, and $\beta_2 = -1.0\pm0.1$. 
At $\Delta t\approx 305$\,days, the 1--4\,GHz band probably remains below the broken frequency, and the blue data in  Figure~\ref{fig:radio_sed} suggests $\beta_1 > 1.7$. 
At $\Delta t \approx 417$\,days, $\nu_{\rm peak} = 4.82_{-1.18}^{+1.36}$\,GHz, $L_{\nu\,\rm peak} = 4.33_{-0.34}^{+0.36}\times 10^{28}\,{\rm erg\,s^{-1}\,Hz^{-1}}$, $\beta_1 = 0.4_{-0.2}^{+0.3}$, and $\beta_2 = -0.5\pm0.1$. Equation~(\ref{eq:GSbrokenPL}) does not provide a decent description for the data.

The radio observations will further be discussed in \S\ref{subsec:shock_CSM}

\subsection{The Host Galaxy}
\subsubsection{Observations}

\begin{figure*}[htbp!]
\centering
\includegraphics[width=0.95\textwidth]{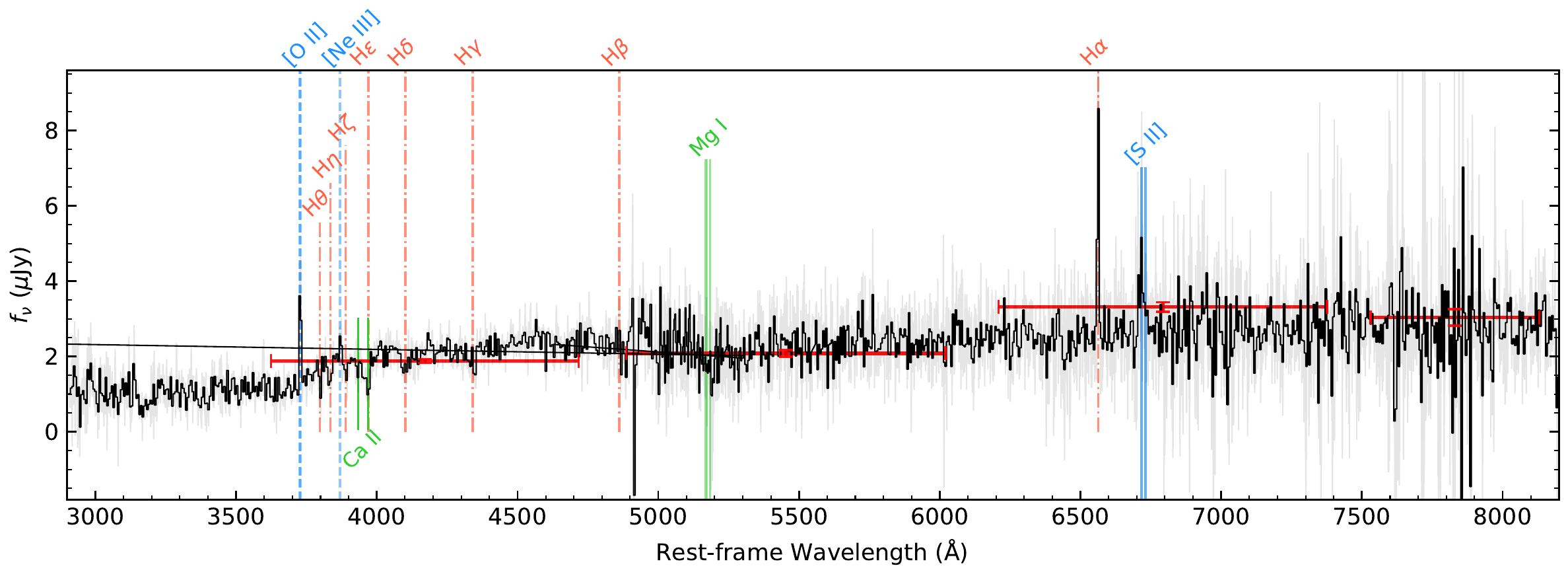}
\includegraphics[width=\textwidth]{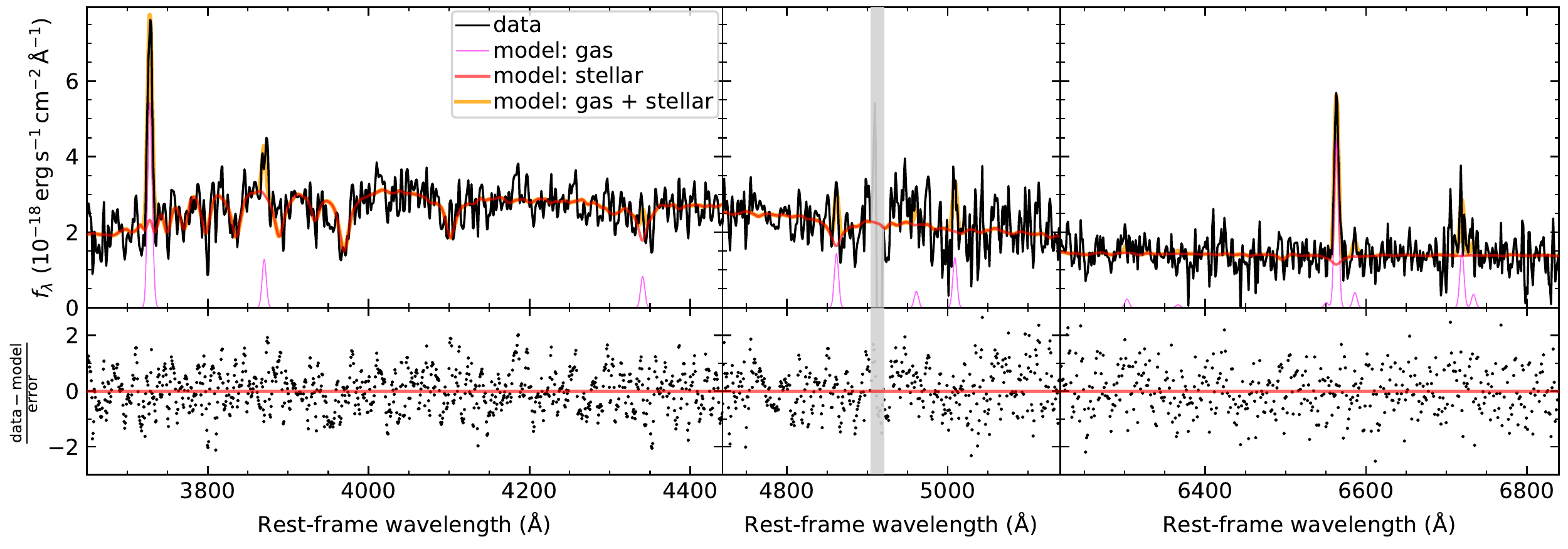}
\caption{
\textit{Upper}: Spectrum (corrected for Galactic extinction) of the host galaxy, overplotted with HSC $griz$ photometry. 
\ad{\textit{Bottom}: Zoomed-in regions of the spectrum (black lines, corrected for Galactic extinction). The thick orange lines show the best-fit \texttt{ppxf} model, which is a combination of the stellar continuum (red lines) and emission lines (thin magenta lines). 
The rest-frame wavelength range 4904--4920\,\AA\ is masked due to the large uncertainty of $f_{\lambda}$ (comtamination by a sky line). }
\label{fig:host}}
\end{figure*}

Deep pre-explosion images of the target field are available in the Hyper Suprime-Cam Subaru Strategic Program (HSC-SSP; \citealt{Aihara2018}) second Public Data Release (PDR2; \citealt{Aihara2019}) and the \textit{Galaxy Evolution Explorer} (\galex; \citealt{Martin2005}) UV imaging survey. As is shown in the left panel of Figure~\ref{fig:host}, \target is $0.50^{\prime\prime}$ offset from an extended blue source  
(${\rm R.A.}=15^{\rm h}47^{\rm m}54.20^{\rm s}$, 
${\rm decl.}=+44^{\circ}39^{\prime}07.01^{\prime\prime}$), which is considered to be the host galaxy. 
\ad{At the host redshift, the spacial offset corresponds to a physical distance of 1.19\,kpc.}
The photometry of the host is shown in Table~\ref{tab:host}.

\begin{deluxetable}{cccc}[htbp!]
    \tablecaption{Observed photometry of the host galaxy.\label{tab:host}}
	\tablehead{
	\colhead{Instrument}
		& \colhead{Band} 
		& \colhead{$\lambda_{\rm eff}$ (\AA)}
		& \colhead{Magnitude} 
	}
	\startdata
\galex & FUV & 1528 & $>23.276$ \\
\galex & NUV & 2271 & $>23.579$ \\
HSC & $g$ & 4755 & $23.282 \pm 0.029 $ \\
HSC & $r$ & 6184 & $23.152 \pm 0.046 $ \\
HSC & $i$ & 7661 & $22.635 \pm 0.040 $ \\
HSC & $z$ & 8897 & $22.721 \pm 0.079 $ \\
HSC & $y$ & 9762 & $22.359 \pm 0.133 $ \\
	\enddata
	\tablecomments{The HSC Kron radius is $0.705^{\prime\prime}$. \galex upper limits are given in $3\sigma$.}
\end{deluxetable}


On 2021 April 14 ($\Delta t = 267.0$\,days), we obtained a spectrum of the host galaxy using the Low Resolution Imaging Spectrometer (LRIS; \citealt{Oke1995}) on the Keck I 10\,m telescope. \ad{We used the 560 dichroic, the 400/3400 grism on the blue side, the 400/8500 grating on the red side, and the $1^{\prime\prime}$ slit width. This setup gives 
a full-width half maximum (FWHM) of $\approx6.8$\,\AA.} Exposure times were 3650 and 3400\,s for the blue and red cameras, respectively. The spectrum (upper panel of Figure~\ref{fig:host}) was reduced and extracted using \texttt{LPipe} \citep{Perley2019lpipe}. 

\subsubsection{Analysis}\label{subsubsec:host_analysis}

In order to determine the redshift and emission line fluxes of the host, we fit the Galactic extinction corrected LRIS spectrum with stellar population models using the penalized pixel-fitting (\texttt{ppxf}) software \citep{Cappellari2004, Cappellari2017}. 
We use the MILES library (${\rm FWHM=2.5\,{\rm \AA}}$; \citealt{Falcon-Barroso2011}), and commonly observed galaxy emission lines, \ad{including H$\alpha$, H$\beta$, H$\gamma$, [\ion{O}{II}], [\ion{S}{II}], [\ion{O}{III}], [\ion{O}{I}], and [\ion{N}{II}]. }
The 
[\ion{O}{I}] $\lambda \lambda 6300$, 6364, 
[\ion{O}{III}] $\lambda \lambda 4959$, 5007 and 
[\ion{N}{II}] $\lambda \lambda 6548, 6583$ 
doublets are fixed at the theoretical flux ratio of 3. 

The best-fit model suggests a redshift of $z=0.1353\pm0.0002$. 
\ad{Zoom-in portions of the spectrum around regions of emission lines are shown in the bottom panel of Figure~\ref{fig:host}.} 
The line fluxes are presented in Table~\ref{tab:host_emi}. 
\ad{Note that since the [\ion{O}{II}] doublets are not resolved, the derived individual line fluxes are not reliable, and we only report the total flux of the doublets.}

\begin{deluxetable}{cc}[htbp!]
	\tablecaption{Galactic extinction corrected galaxy emission line fluxes.\label{tab:host_emi}}
	\tablehead{
		\colhead{Line}  
		& \colhead{Flux ($10^{-18}\,{\rm erg\,s^{-1}\,cm^{-2}\,\AA^{-1}}$)}
	}
	\startdata
	\ad{[\ion{O}{II}] $\lambda\lambda3726$, 3729} & \ad{$53.00\pm 6.09$}   \\\relax
	[\ion{Ne}{III}] $\lambda3869$  & \ad{$10.96 \pm 1.97$} \\ \relax 
	H$\gamma$ $\lambda 4340$ & \ad{$6.31\pm 1.55$} \\\relax
	H$\beta$ $\lambda 4861 $ & \ad{$9.81 \pm 1.87$} \\\relax
	[\ion{O}{III}] $\lambda \lambda 4959$, 5007 & \ad{$11.67\pm 4.57$ (2.6$\sigma$)} \\\relax 
	[\ion{O}{I}] $\lambda \lambda 6300$, 6364  & \ad{$1.59\pm1.17$ (1.4$\sigma$)} \\\relax 
    H$\alpha$ $\lambda 6563$ & \ad{$22.82\pm 0.89$} \\\relax
    [\ion{N}{II}] $\lambda \lambda 6548, 6583$  &  \ad{$2.72\pm 1.71$ (1.6$\sigma$)} \\\relax 
	[\ion{S}{II}] $\lambda6716$ & \ad{$7.32 \pm 2.01$} \\\relax 
	[\ion{S}{II}] $\lambda6731$ & \ad{$1.77 \pm 1.26$ (1.4$\sigma$)} \\
\enddata
\tablecomments{\ad{Marginally detected emission lines are indicated with the detection significance shown in the parenthesis. }}
\end{deluxetable}

\begin{deluxetable}{cc}[htbp!]
	\tablecaption{Emission line ratios.\label{tab:line_ratio}}
	\tablehead{
		\colhead{Definition}  
		& \colhead{Value}
	}
	\startdata
	\ad{[\ion{O}{III}]$\lambda$5007/H$\beta$ } & \ad{$0.90^{+0.78}_{-0.58}$ }\\\relax
	\ad{$\rm log$\{[\ion{O}{III}]$\lambda$5007/H$\beta$\} }& \ad{$-0.05^{+0.27}_{-0.45}$} \\\relax
	\ad{[\ion{N}{II}]$\lambda6583$/H$\alpha$ }& \ad{$<0.18$ }\\\relax
	\ad{$\rm N2 \equiv log$\{[\ion{N}{II}]$\lambda6583$/H$\alpha$\} }& \ad{$<-0.73$} \\\relax
	\ad{$\rm O3N2 \equiv log$\{[\ion{O}{III}]$\lambda$5007/H$\beta$\}$-$N2 }& \ad{$<-1.71$} \\\relax
	\ad{[\ion{S}{II}]$\lambda\lambda6716$,31/H$\alpha$ }& \ad{ $0.40\pm0.17$ } \\\relax
	\ad{log\{[\ion{S}{II}]$\lambda\lambda6716$,31/H$\alpha$\} }& \ad{ $-0.40^{+0.16}_{-0.24}$} \\\relax
	\ad{[\ion{O}{I}]$\lambda6300$/H$\alpha$} & \ad{ $<0.12$ } \\\relax
	\ad{log\{[\ion{O}{I}]$\lambda6300$/H$\alpha$\} } & \ad{ $<-0.94$  }
\enddata
\tablecomments{\ad{Line ratios and their uncertainties are estimated using the 5$^{\rm th}$, 50$^{\rm th}$ and 95$^{\rm th}$ percentiles of the MC simulations. When the 5$^{\rm th}$ percentile value is negative, we present the 95$^{\rm th}$ percentile as an upper limit. }}
\end{deluxetable}

\begin{figure}[htbp!]
    \centering
    \includegraphics[width=\columnwidth]{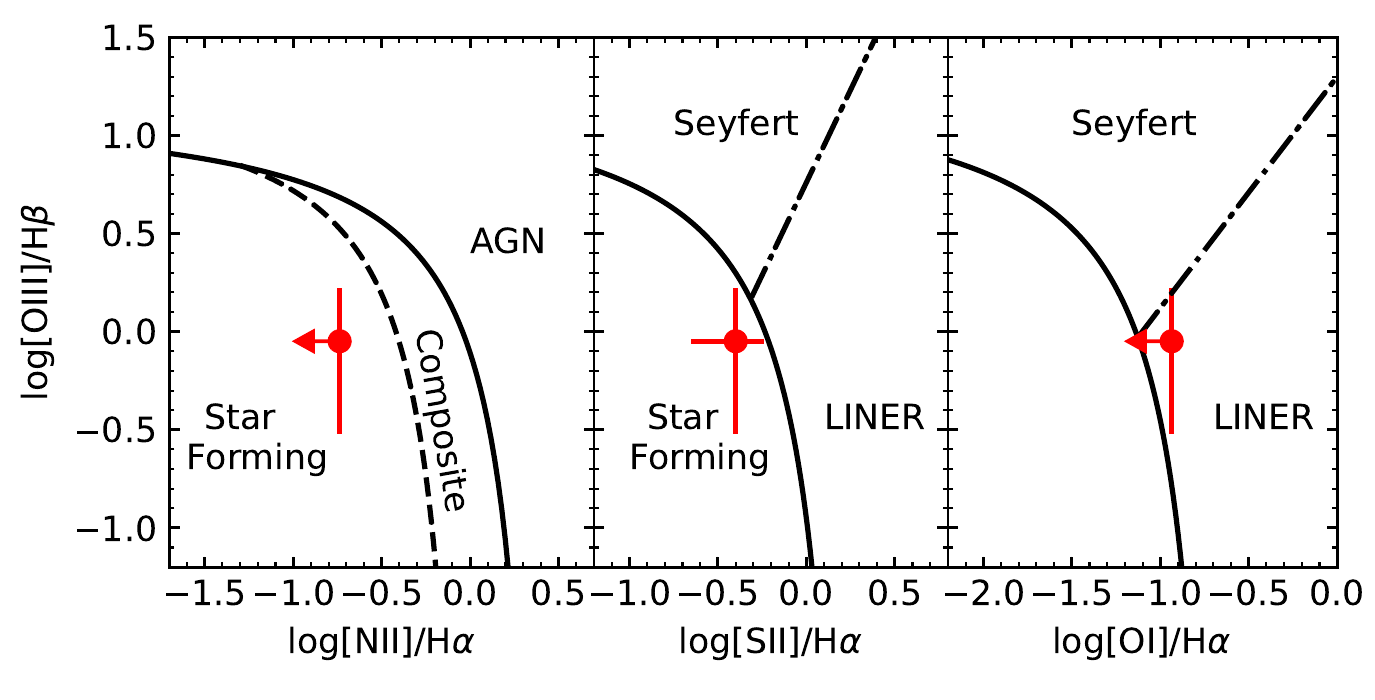}
    \caption{The host galaxy of \target on the BPT diagrams. The diagnostic lines follow Fig.~4 of \citet{Kewley2006}. \label{fig:BPT}}
\end{figure}

\ad{
The calculated line ratios are given in Table~\ref{tab:line_ratio}. 
Uncertainties in line ratios are calculated by performing $10^4$ Monte Carlo (MC) trials using the measured flux uncertainties. 
Figure~\ref{fig:BPT} shows the location of the host galaxy on the Baldwin, Phillips, \& Terlevich (BPT) diagrams \citep{Baldwin1981}. Under the diagnostic definitions of \citet{Kewley2006}, the host falls in the region of star-forming galaxies.
}

\ad{We measure the oxygen abundance using two metallicity indicators N2 and O3N2 \citep{Pettini2004}, defined in Table~\ref{tab:line_ratio}. Using the calibration reported by \citet{Marino2013},
the gas-phase oxygen abundance is 
$<8.40\pm0.16({\rm sys})$ in the N2 scale, and 
$>8.17\pm0.18({\rm sys})$ in the O3N2 scale. }
Compared with the solar metallicity ($Z_\odot$) of $12 + {\rm log}({\rm O/H}) = 8.69$ \citep{Asplund2009}, 
\ad{our constraints suggest a metallicity of $10^{-0.70}$--$10^{-0.13} Z_\odot$.}

To obtain an estimate of the host galaxy total stellar mass ($M_\ast$), we fit the host SED with flexible stellar population synthesis (FSPS; \citealt{Conroy2009}) models \citep{Foreman-Mackey2014}. We adopt a delayed exponentially declining star-formation history (SFH) characterized by the $e$-folding timescale $\tau_{\rm SFH}$, such that the time-dependent star-formation rate $\psi_\ast(t)\propto t e^{-(t/\tau_{\rm SFH})}$. 
The \texttt{Prospector} package \citep{Johnson2021} was used to run a Markov Chain Monte Carlo (MCMC) sampler \citep{Foreman-Mackey2013}. 
We use log-uniform priors for the following three parameters: 
$M_\ast$ in the range [$10^7\,M_\odot$, $10^9\,M_\odot$], 
$\tau_{\rm SFH}$ in the range [0.1\,Gyr, 100\,Gyr], 
the metallicity \ad{${{\rm log}(Z/Z_\odot)}$ in the range $-0.70$ and $-0.13$,}
and the population age $t_{\rm age}$ in the range [0.1\,Gyr, 12.5\,Gyr]. 
Host galaxy extinction was included, with $E(B-V)_{\rm host}$ uniformly sampled between 0 and 1.  
From the marginalized posterior probability functions we obtain 
\ad{${\rm log}(M_\ast/M_\odot) = 7.94_{-0.39}^{+0.22}$, 
${\rm log}(Z/Z_\odot) = -0.46_{-0.17}^{+0.20}$, 
$\tau_{\rm SFH} = 11.6_{-10.0}^{+45.6}$\,Gyr, 
$t_{\rm age} = 1.82_{-1.50}^{+4.07}$\,Gyr, and 
$E(B-V)_{\rm host} = 0.21_{-0.12}^{+0.10}$,
where uncertainties are epresented by the 68\% confidence intervals.}

\ad{
Using the 90\% confidence interval of the $M_\ast$ posterior probability function and the mass--metallicity relation (MZR) of low-mass galaxies \citep{Berg2012}, we infer that the typical log($Z/Z_\odot$) at the host mass should be $-0.78^{+0.10}_{-0.16}\pm0.15({\rm sys})$. 
The measured metallicity is therefore on the high end of the distribution.
}

We convolve the observed LRIS spectrum with the HSC $i$-band filter and compare the flux with the host photometry (Table~\ref{tab:host}), which suggests that 80.6\% of the total host flux is captured by the LRIS slit. Subsequently, we assume the same fraction of total H$\alpha$ flux is captured by the slit and no host extinction, and calculate the H$\alpha$ luminosity to be \ad{$L_{\rm H\alpha} = (1.39\pm0.05) \times10^{39}\,{\rm erg\,s^{-1}}$}. Using the \citet{Kennicutt1998} relation converted to a Chabrier initial mass function \citep{Chabrier2003, Madau2014}, we infer a star formation rate (SFR) of 
\ad{$(6.93\pm0.27)\times 10^{-3} \,M_\odot \,{\rm yr^{-1}}$}. 
An extinction of \ad{$E(B-V)_{\rm host}\sim0.21$} will render the SFR higher by a factor of \ad{$\sim1.5$}.
\ad{Therefore, hereafter we adopt ${\rm SFR} = 6.93^{+3.90}_{-0.27} \times 10^{-3} \,M_\odot \,{\rm yr^{-1}}$.}
\ad{The specific star formation rate is ${\rm sSFR}\equiv {\rm SFR}/M_\ast = 0.80^{+0.45}_{-0.03}\times10^{-10}\,{\rm yr^{-1}}$, where we only consider the uncertainty of SFR but exclude the uncertainty of $M_\ast$. } 

\section{Inferences and Discussion} \label{sec:discussion} 

\subsection{A Mildly Relativistic Shock in a Dense Environment}  \label{subsec:shock_CSM}
\subsubsection{Standard SSA Modeling} \label{subsubsec:C98SSA}
At $\Delta t \approx 261$\,days, the observed spectral index of $\beta_2 \approx -1$ (\S\ref{subsec:radio_obs}) in the optically thin regime of the radio SED motivates us to adopt the standard model given by \citet{Chevalier1998}, where the electrons in the CSM are accelerated by the forward shock into a power-law distribution of energy $N(E) = N_0 E^{-p}$. We do not consider the alternative of a relativistic Maxwellian electron-energy distribution, in which case we expect a much steeper $\beta_2$ (see, e.g., Fig.~11 of \citealt{Ho2021b}) and a shock speed of $v_{\rm sh}\gtrsim 0.2c$ \citep{Margalit2021_sync}. The $v_{\rm sh}$ inferred from our observations is much slower (see below). We note that the standard model might not be fully appropriate since the observed spectral index of $\beta_1$ in the optically thick regime is much shallower than the $\beta_1 =2.5$ expected from SSA. We investigate the effects of CSM inhomogeneity and scintillation in \S\ref{subsubsec:CSMinhomo}.

In the standard model of \citet{Chevalier1998}, the minimum electron energy is $E_{\rm min}=511$\,keV; the peak of the SED is governed by synchrotron self-absorption (SSA) such that $\tau(\nu_{\rm peak}) = 1$; the radio emitting region is approximated by a sphere with radius $R$ and volume filling factor $f$ (hereafter assumed to be 0.5); the magnetic energy density ($U_B\propto B^2$) and the relativistic electron energy density ($U_e\propto N_0$) are assumed to scale as the total (thermalized) post-shock energy density $U$, such that $U_B = \epsilon_B U$ and $U_e = \epsilon_e U$. 

We define $L_{\theta\nu}\equiv   4\pi D_{\theta}^2 f_\nu= L_{\nu} / (1+z)^4$, $L_{\theta\nu,29}\equiv L_{\theta\nu, \rm peak}/(10^{29}\,{\rm erg\,s^{-1}\,Hz^{-1}})$ and $\nu_5 \equiv \nu_{\rm peak} / (5\,\rm GHz)$, such that
\begin{subequations}
\begin{align}
    R =& 7.1\times 10^{16} \left( \frac{\epsilon_e}{\epsilon_B} \right)^{-1/19} L_{\theta\nu,29}^{9/19} \nu_5^{-1}\,{\rm cm}\\
    B =& 0.36 \left(\frac{\epsilon_e}{\epsilon_B} \right)^{-4/19} L_{\theta\nu,29}^{-2/19} \nu_5\,{\rm G} \\
   U = & 4.0\times 10^{48} \frac{1}{\epsilon_B} \left(\frac{\epsilon_e}{\epsilon_B}\right)^{-11/19}  L_{\theta\nu,29}^{23/19} \nu_5^{-1}\,{\rm erg}. \label{eq:U}
\end{align}
\end{subequations}
The upstream CSM density can be estimated under the conditions of strong shocks and fully ionized hydrogen (see Eq.~16 of \citealt{Ho2019}):
\begin{align}
    n_e =& 61  \frac{1}{\epsilon_B} \left( \frac{\epsilon_e}{\epsilon_B} \right)^{-6/19} L_{\theta\nu,29}^{-22/19}  \nu_5^{4} \left(\frac{\Delta t}{\rm 100\,days}\right)^2  \,{\rm cm^{-3}}
\end{align}
Assuming that the CSM density profile is determined by a pre-explosion steady wind with mass-loss rate $\dot M$ and velocity $v_{\rm w}$, we have (see Eq.~23\footnote{The normalization constant in Eq.~23 of \citealt{Ho2019} is off by a factor of $\sim10$. Here we update the equation with the correct constant.} of \citealt{Ho2019}):
\begin{align}
    \frac{\dot M}{v_{\rm w}} \left( \frac{1000\,{\rm km\,s^{-1}}}{10^{-4}\,M_\odot\,{\rm yr^{-1}}}\right) =& 0.10 \left(\frac{1}{\epsilon_B} \right) \left( \frac{\epsilon_e}{\epsilon_B} \right)^{-8/19}  \notag \\
    & \times L_{\theta\nu, 29}^{-4/19} \nu_5^{2} \left( \frac{\Delta t}{100\,{\rm days}}\right)^{2}
\end{align}

We adopt $L_{\theta\nu\,{\rm peak}} = L_{\nu\,{\rm peak}} / (1+z)^4 \approx 1.0\times 10^{29}\,{\rm erg\,s^{-1}\,Hz^{-1}}$ and $\nu_{\rm peak}\approx 7$\,GHz at $\Delta t=261$\,days.
Assuming $\epsilon_e = \epsilon_B = 1/3$, we have 
$R \approx 5.1 \times 10^{16}\,{\rm cm}$, 
$B \approx 0.50\,{\rm G}$, 
$U \approx 1.7 \times 10^{49}\,{\rm erg}$, 
and $n_e \approx 3.5\times 10^3\,{\rm cm^{-3}}$. 
Assuming $\epsilon_e = 0.1$, $\epsilon_B = 0.01$, we have 
$R \approx 4.6 \times 10^{16}\,{\rm cm}$, 
$B \approx 0.31\,{\rm G}$, 
$U \approx 1.5 \times 10^{50}\,{\rm erg}$, 
and $n_e \approx 5.6\times 10^4\,{\rm cm^{-3}}$. 
The average shock velocity ($v_{\rm sh} = R / \Delta t$) is 0.07--0.08$c$, suggesting a mildly relativistic shock. 
The derived \ad{$R$, $U$, $v_{\rm sh}$ should be taken as upper limits, $B$, $n_e$, $\dot M / v_{\rm w}$ should be taken as upper limits.} See the discussion in \S\ref{subsubsec:CSMinhomo}.

\begin{figure}[htbp!]
    \centering
    \includegraphics[width = \columnwidth]{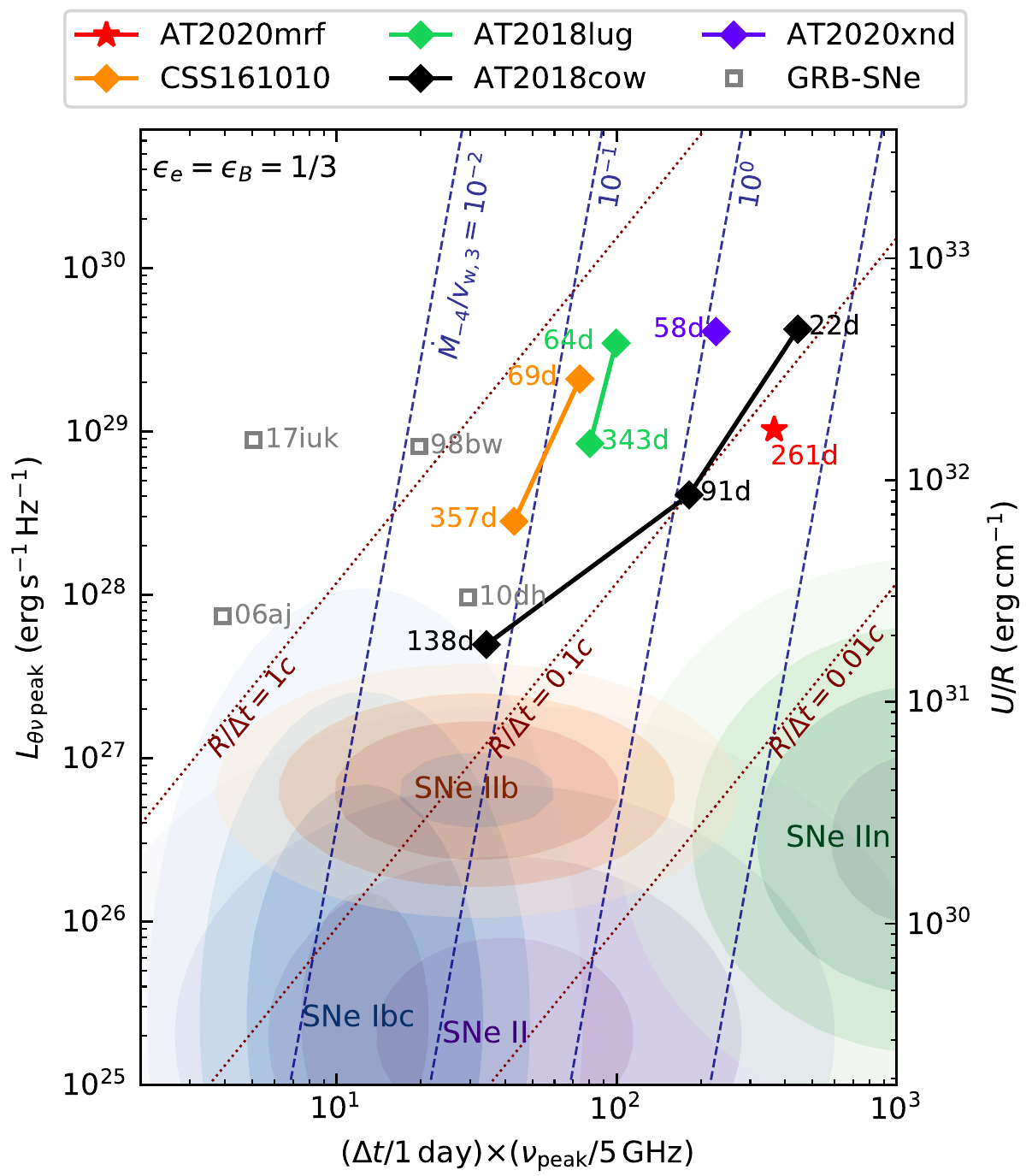}
    \includegraphics[width = \columnwidth]{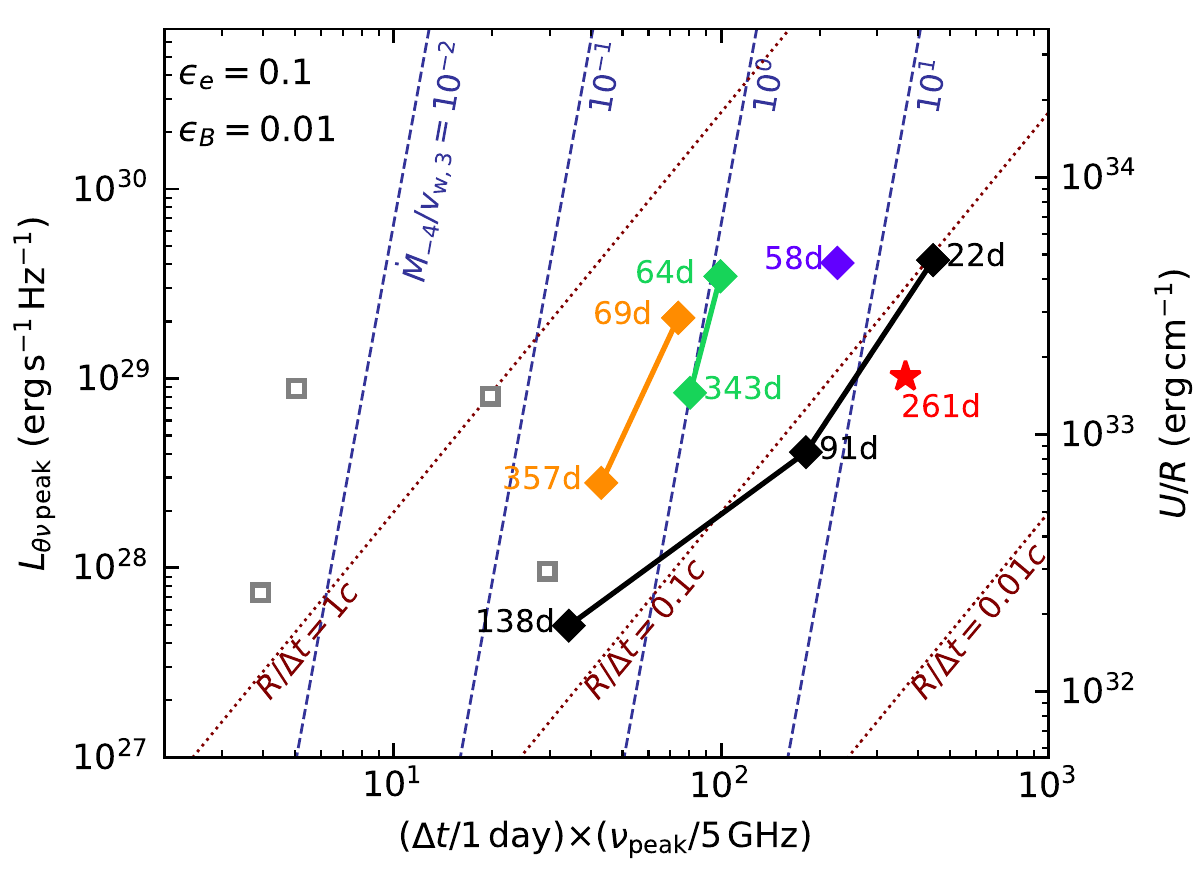}
    \caption{Peak radio luminosity ($L_{\nu\,{\rm peak}}$) versus the product of peak time and $\nu_{\rm peak}$.
    Under the assumptions of the standard SSA model, dotted lines mark constant time-averaged velocity; 
    Dashed lines mark constant mass-loss rate ($\dot M_{-4}\equiv \dot M / (10^{-4}\, M_{\odot}\,{\rm yr^{-1}})$) scaled to wind velocity ($v_{\rm w, 3}\equiv v_{\rm w} / (10^3\,{\rm km\,s^{-1}})$).
    The two panels show the results with different assumptions of $\epsilon_e$ and $\epsilon_B$. 
 The data of AT2018cow-like objects and GRB-SNe are based on Fig.~9 of \citealt{Ho2021b} and Fig.~3 of \citet{Nayana2021}. 
    \label{fig:shock}}
\end{figure}

The upper panel of Figure~\ref{fig:shock} compares \target with normal SNe \citep{Bietenholz2021}, SNe associated with long GRBs, and four AT2018cow-like objects in the literature. Note that all GRB-SNe are of type Ic-BL. 
The peak luminosity of \target is much greater than normal SNe and is in the same regime as other AT2018cow-like objects. A physical interpretation is that the energy divided by the shock radius ($U/R \propto L_{\theta\nu\,{\rm peak}}^{14/19}$) is greater.
This indicates a more efficient conversion/thermalization of energy, which can come from a higher explosion energy or a higher ambient density \citep{Ho2019}. 

Moreover, we see that the CSM ``surface density'' ($n_e R^2 \propto \dot M / v_{\rm w}$) of \target at 261\,days is similar to AT2018cow at 22\,days. 
At a similar shock radius of $R \sim 6 \times 10^{16}$\,cm, the CSM number density of AT2018cow is $n_e < 33\,{\rm cm^{-3}}$ \citep{Nayana2021} --- more than 100 times smaller than that in AT2020mrf. 
Since $\dot M / v_{\rm w}$ generally decreases at later times (i.e., the density profile is steeper than $n_e \propto r^{-2}$), the immediate environment of \target is probably denser than all other AT2018cow-like events.

\subsubsection{CSM Inhomogeneity and Scintillation} \label{subsubsec:CSMinhomo}

\begin{figure}[htbp!]
    \centering
    \includegraphics[width = \columnwidth]{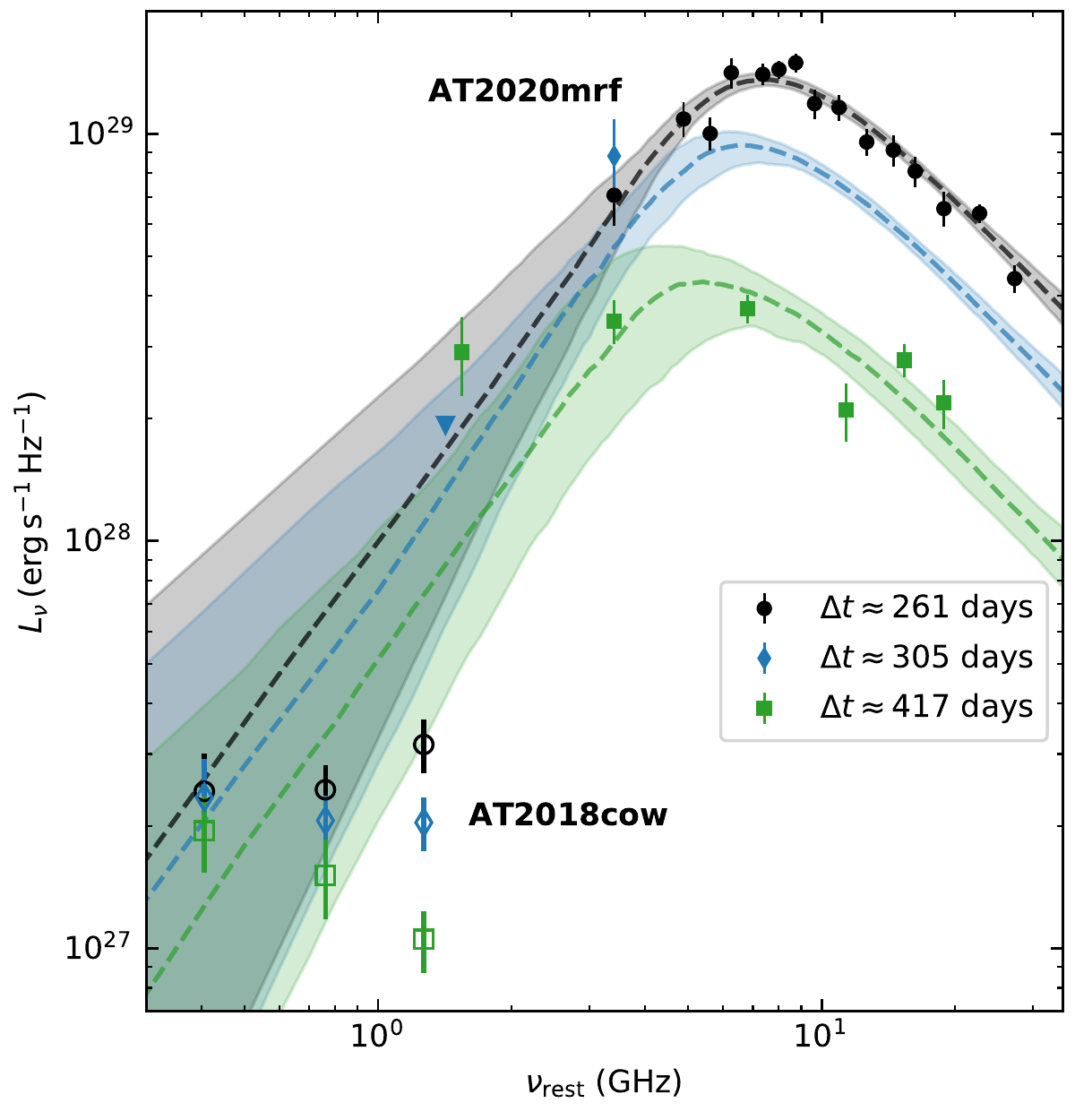}
    \caption{Radio SEDs of \target (solid markers), compared with the that of AT2018cow at similar phases (hollow markers, interpolated from Table~1 of \citealt{Nayana2021}). Dashed lines are the inhomogeneous SSA model fits to the observations of \target. \label{fig:radio_sed_inhomo}}
\end{figure}

\begin{figure*}[htbp!]
    \centering
    \includegraphics[width=0.7\textwidth]{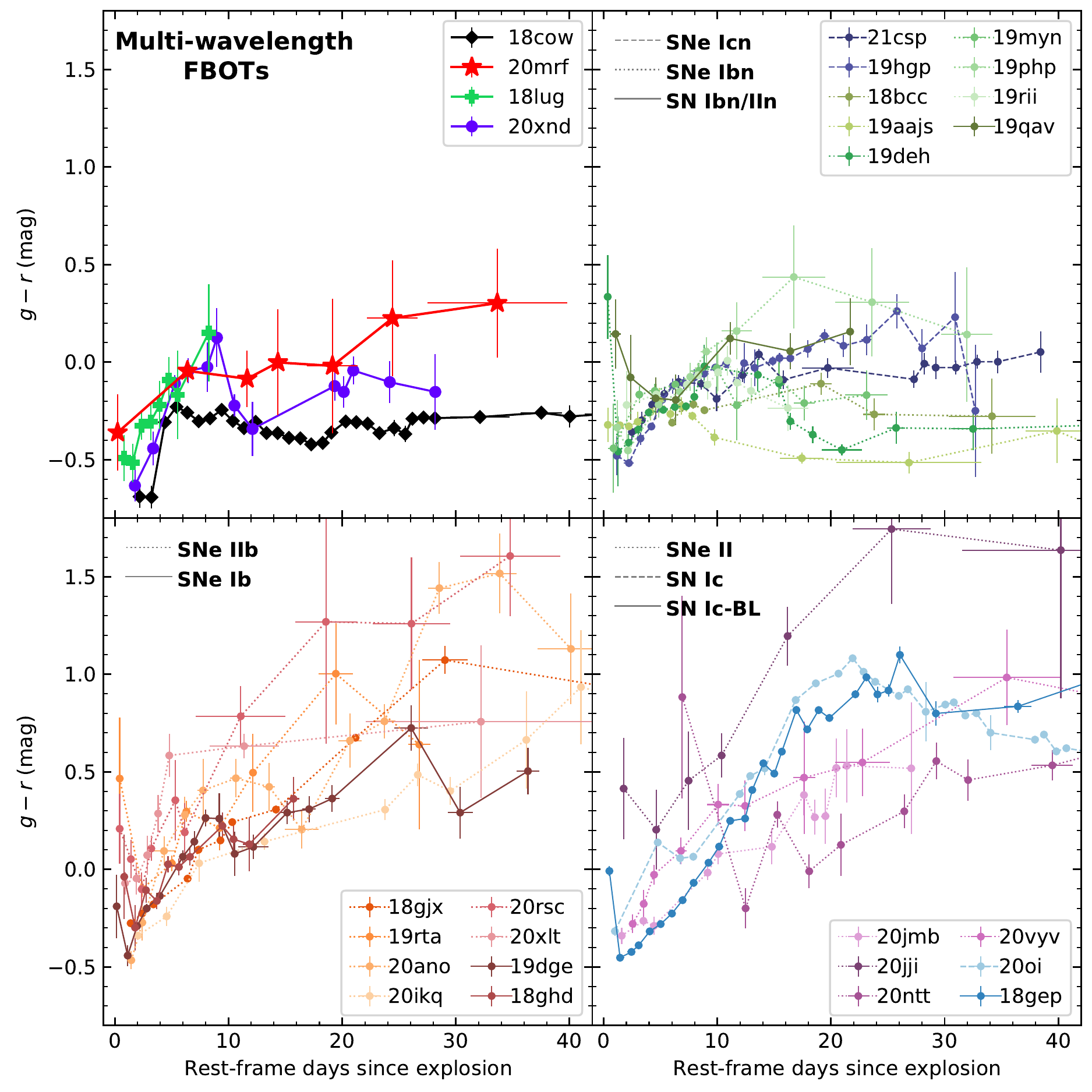}
    \caption{Color evolution of FBOTs. The upper left panel shows four events associated with bright radio emission. The upper right panel shows interacting SNe of type Icn, Ibn, and IIn. The lower panels show type II SNe, as well as stripped envelope SNe of type IIb, Ib, Ic, and Ic-BL. 
    \label{fig:color_evol}}
\end{figure*}

The small values of $\beta_1$ and the flat-topped radio SEDs (Figure~\ref{fig:radio_sed}) motivate us to assume an inhomogeneous CSM, which means that the distribution of electrons or magnetic field strength varies within the synchrotron source \citep{Bjornsson2017}. In this model, 
between the standard SSA optically thick $F_\nu \propto \nu^{5/2}$ regime and the optically thin $F_\nu \propto \nu^{-(p+1)/2}$ regimes, there is a transition regime with a spectral index of $0<\beta<2.5$. Since the measured $\beta_1$ remains below $2.5$, we assume that the standard SSA optically thick regime is at frequencies lower than our observations. 

Following \citet{Chandra2019}, we fit the full set of radio data with the function
\begin{align}
    L(\nu, t) = K_1 \nu_5^{\beta} \left(\frac{\Delta t}{100\,{\rm days}}\right)^{a} \left[ 1 - {\rm exp}\left (  - \tau_{\rm ssa}(\nu, t)\right)\right],
\end{align}
where $\tau_{\rm ssa}$ is the SSA optical depth
\begin{align}
    \tau_{\rm ssa}(\nu, t) =  K_2 \nu_5^{-(\beta + \frac{p-1}{2})} \left(\frac{\Delta t}{100\,{\rm days}}\right)^{-(a + b)}.
\end{align}

The best-fit model is shown in Figure~\ref{fig:radio_sed_inhomo}, with $K_1 = 5.4_{-4.4}^{+14.6}\times 10^{29}\,{\rm erg\,s^{-1}\,Hz^{-1}}$, $K_2 = 13_{-10}^{+53}$, $\beta = 1.6_{-0.6}^{+0.8}$, $p = 3.3_{-0.3}^{+0.4}$, $a = -1.6_{-1.4}^{+1.8}$, and $b = 3.0_{-0.4}^{+0.3}$. Evidence of source inhomogeneities has also been found in AT2018cow \citep{Nayana2021}. With an inhomogeneous CSM, the $R$, $U$, $v_{\rm sh}$ \ad{values derived in \S\ref{subsubsec:C98SSA} should be taken as lower limits, and $B$, $n_e$, $\dot M / v_{\rm w}$ should be taken as upper limits.}

A few datapoints at $\Delta t > 300$\,days are not well fitted by the inhomogeneous SSA model. 
We estimate the effects of interstellar scintillation (ISS) to our radio observations using the \texttt{NE2001} model \citep{Cordes2002} of the Galactic distribution of free electrons. The transition frequency below which strong scattering occurs is \citep{Goodman1997}:
\begin{align}
    \nu_{\rm ss} = 10.4 ({\rm SM_{-3.5}})^{6/17} d_{\rm scr, kpc}^{5/17}\,{\rm GHz},
\end{align}
where ${\rm SM}_{-3.5} \equiv {\rm SM} /( 10^{-3.5} \,{\rm m}^{-20/3}\,{\rm kpc})$ is the scintillation measure, and  $d_{\rm scr, kpc}$ is the distance to the electron scattering screen in kpc.
For the line of sight to \target (Galactic coordinates $l = 71.339^{\circ}$, $b = 50.806^{\circ}$), \texttt{NE2001} predicts $\nu_{\rm ss} = 8.3$\,GHz
and ${\rm SM}_{-3.5}  =  0.53$, implying $d_{\rm scr, kpc} = 1.0$. 
This suggests that the 11.35\,GHz ``dip'' (or 15--19\,GHz ``excess'') cannot be explained by ISS.

AT2020mrf is subject to diffractive or refractive ISS if the source angular size satisfies $\theta_{\rm s} < 3.3 \nu_{10}^{6/5}\,{\rm \mu as}$ or $\theta_{\rm s} < 2.0 \nu_{10}^{-11/5}\,{\rm \mu as}$ \citep{Goodman1997}.
We have shown that the shock radius at times of our radio observations is $R\gtrsim 5\times 10^{16}$\,cm, corresponding to $\theta_{\rm s} \gtrsim 6.8\,\mu{\rm as}$. Therefore, the 3.4\,GHz ``excess'' at $\Delta t \approx 305$\,days and the 1.5\,GHz ``excess'' at $\Delta t \approx 417$\,days are likely caused by refractive ISS.

\subsection{Properties of the Optical Emission} \label{subsec:optlc_model}
\subsubsection{Rise and Decline Timescales}
To constrain the optical evolution of \target around maximum, we model the multi-band photometry using a power-law rise and an exponential decay. For simplicity we assume a blackbody SED and a single temperature for data at $\Delta t< 15$\,days. The best-fit model in the $r_{\rm ZTF}$ band is shown as the solid orange line in Figure~\ref{fig:optlc}.

To compare \target with the sample of spectroscopically classified FBOTs presented by \citet{Ho2021a}, we calculate the time it takes for \target to rise from half-max to max ($t_{1/2, \rm rise} = 2.4\pm 0.2$\,days), and to decline from max to half-max ($t_{1/2, \rm fade} = 4.8\pm0.2$\,days). 
Its total duration above half-max is $t_{1/2} = 7.1_{-0.2}^{+0.3}$\,days. On the $M_{\rm peak}$ versus $t_{1/2}$ diagram (see, e.g., Fig.~1 of \citealt{Ho2021a} and Fig.~7 of \citealt{Perley2022}), \target lies between previously studied AT2018cow-like events ($t_{1/2}\lesssim 5$\,days, $M_{\rm peak}\lesssim -20.5$) and interacting SNe of type IIn/Ibn/Icn ($t_{1/2}\gtrsim 7$\,days, $M_{\rm peak}\gtrsim -20.0$).

\subsubsection{Color Evolution}

The $g-r$ color of \target is $-0.34 \pm 0.20$\,mag at the day of discovery ($\Delta t\approx 0.25$\,day), and reddens at later times. At $\Delta t \approx 6.4$, 11.7, and 23--28\,days, the $g-r$ values are $-0.05\pm0.06$\,mag, $-0.09\pm0.14$\,mag, and $0.05 \pm0.27$\,mag, respectively. Assuming that the optical SED can be modeled by a blackbody, the blackbody temperature ($T_{\rm bb}$) decreases from $\sim 2\times 10^4$\,K to $\sim 10^4$\,K. Similar \ad{cooling} signatures have also been observed in AT2018lug, while both AT2018cow and AT2020xnd remain blue post-peak. 

Figure~\ref{fig:color_evol} compares the color evolution of \target with other FBOTs. We have included AT2018cow \citep{Perley2019}, AT2018lug \citep{Ho2020}, AT2020xnd \citep{Perley2021xnd}, the type Icn SNe 2019hgp \citep{Gal-Yam2021hgp} and 2021csp \citep{Perley2022}, as well as the gold sample of 22 spectroscopically classified FBOTs presented by \citet{Ho2021a}. The calculated $g-r$ color has been corrected for Galactic extinction but assumes no host reddening. As can be seen, the amount of \ad{$g-r$ increase} observed in \target is closer to other multi-wavelength FBOTs and interacting SNe, but smaller than events shown in the lower panels.

\subsubsection{Possible Power Sources} \label{subsubsec:opt_power}
Like many other FBOTs, the fast rise and luminous optical peak of \target is unlikely to be powered by radioactive $^{56}$Ni decay, which would require the nickel mass $M_{\rm Ni}$ to be greater than the ejecta mass $M_{\rm ej}$ (see, e.g., Fig.~1 of \citealt{Kasen2017}). Possible emission mechanisms include shock breakout (SBO) from extended CSM \citep{Waxman2017}, shock cooling emission (SCE) from an extended envelope \citep{Piro2021}, \ad{continued interaction between the SN ejecta and the CSM \citep{Smith2017, Fox2019}}, and reprocessing of X-ray/UV photons (potentially deposited by a central engine) by dense outer ejecta \citep{Margutti2019} or an optically think wind \citep{Piro2020}. We do not attempt to distinguish between these scenarios due to a lack of multi-wavelength observations at early times.

\begin{figure}[htbp!]
    \centering
    \includegraphics[width = \columnwidth]{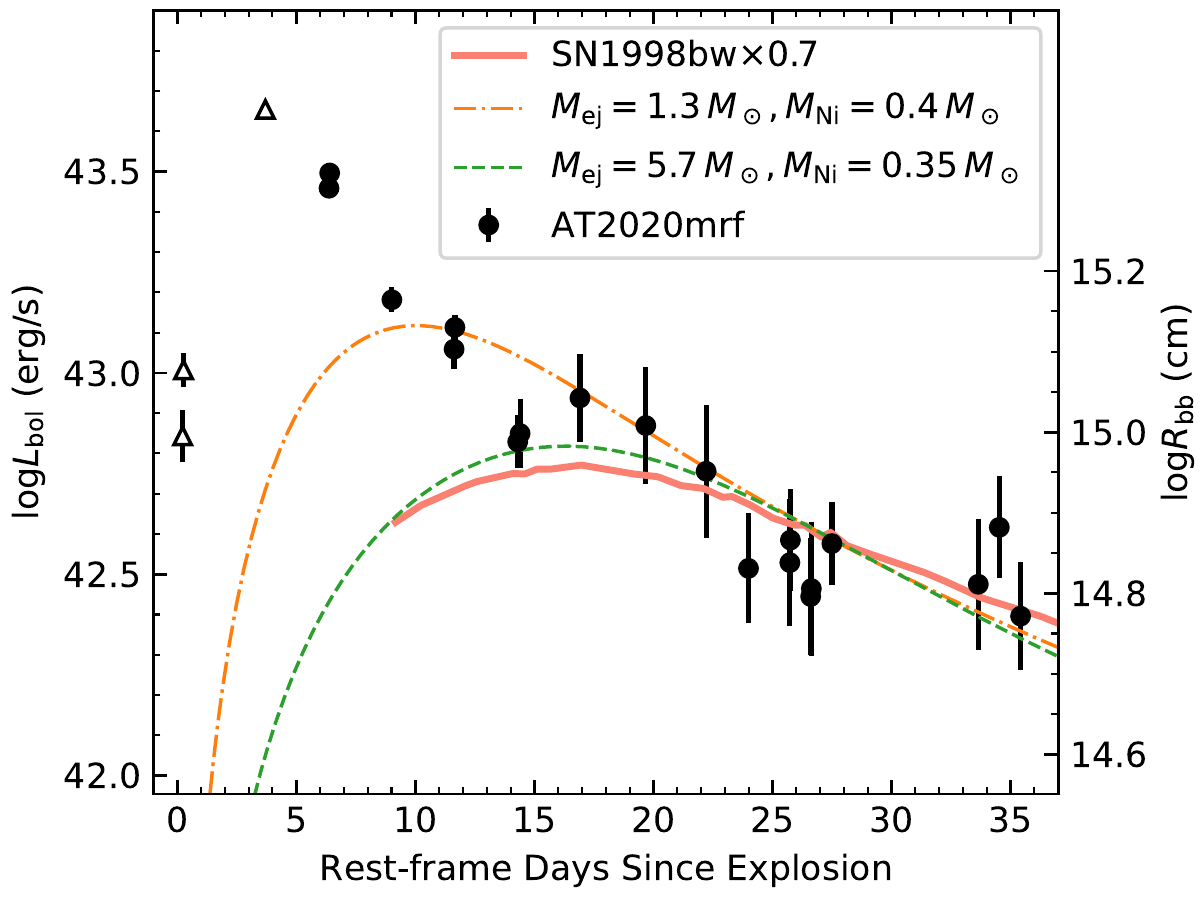}
    \caption{Bolometric light curve of \target converted from ZTF photometry, assuming $T_{\rm bb} = 10^4$\,K. Data at $<5$\,days are shown as upward triangles since the temperature at early time is $>10^4$\,K. The $L_{\rm bol}$ of SN1998bw \citep{Galama1998} is shown for comparison. 
    We show two models of radioactivity powered SN in the photospheric phase \citep{Valenti2008, Lyman2016}, adopting an opacity of $\kappa = 0.07\,{\rm cm^2\,g^{-1}}$ (typical for stripped envelope SNe; \citealt{Taddia2018}), and a photospheric velocity of $v_{\rm phot} = 2\times 10^4\,{\rm km\,s^{-1}}$ (typical for GRB-SNe; \citealt{Modjaz2016}).
    \label{fig:Lbol}}
\end{figure}

\begin{figure*}[htbp!]
    \centering
    \includegraphics[width=0.49\textwidth]{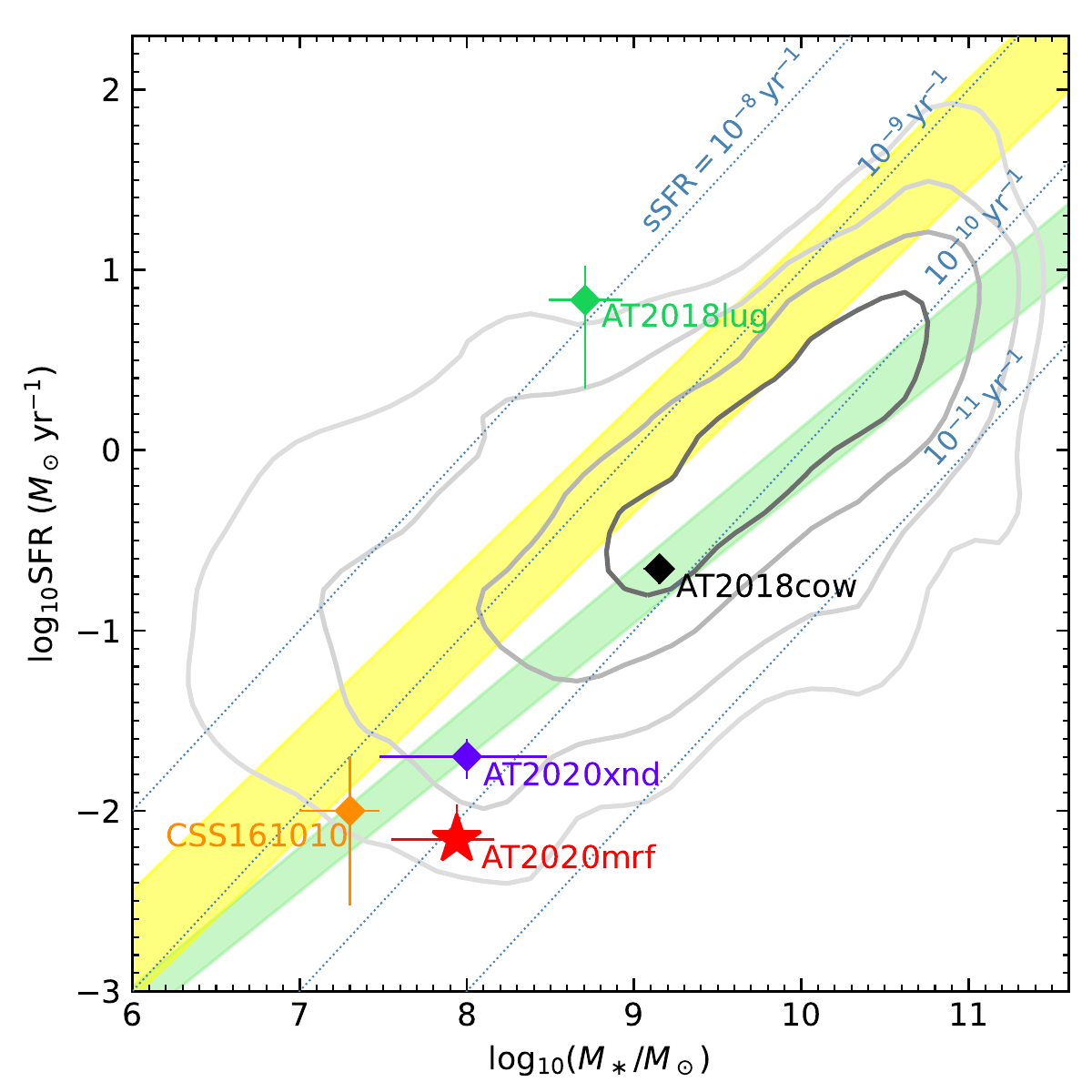}
    \includegraphics[width=0.49\textwidth]{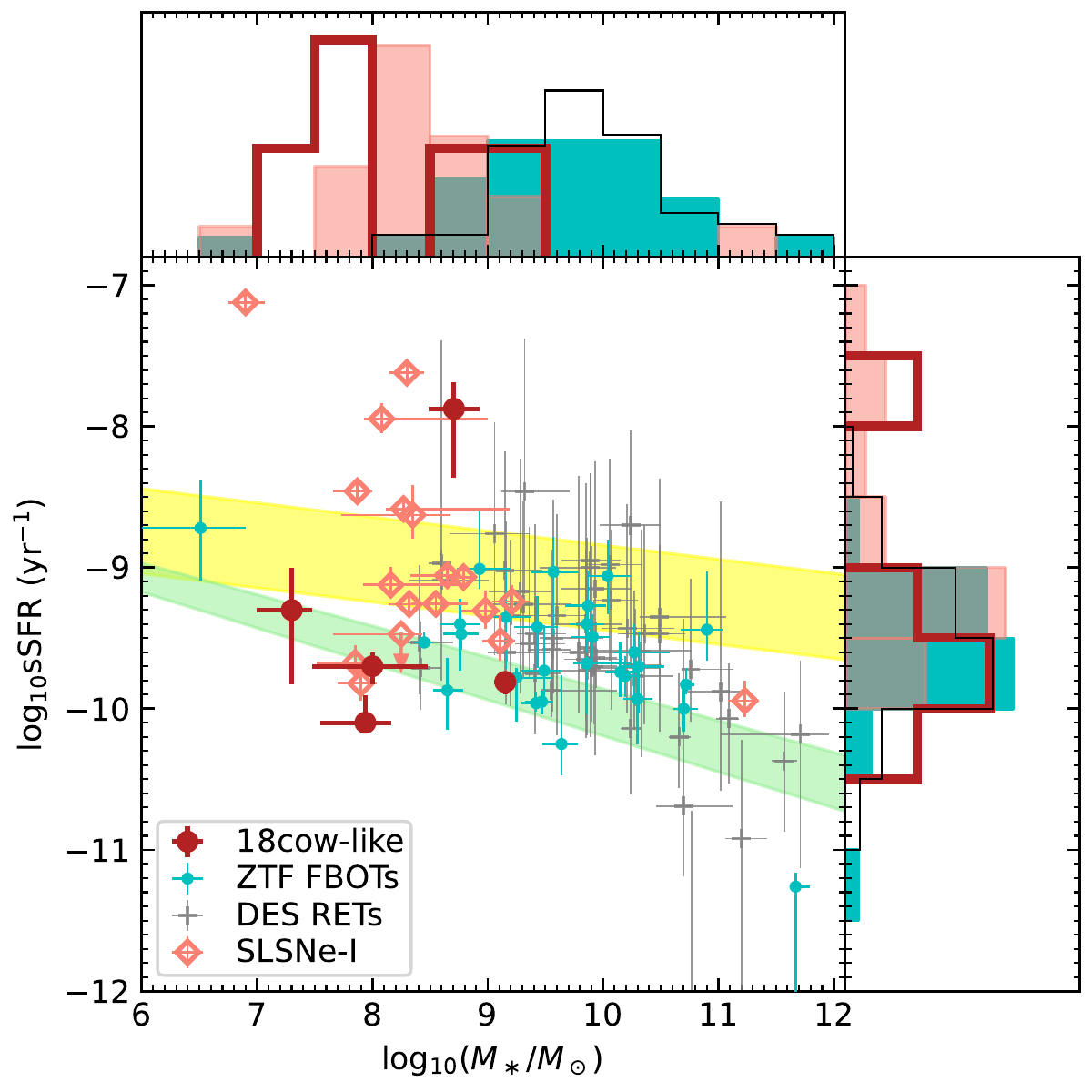}
    \caption{\textit{Left}:
    The host galaxy of \target on the SFR--$M_\ast$ diagram, compared to hosts of other AT2018cow-like events: AT2018cow itself \citep{Perley2019}, AT2018lug \citep{Ho2020}, CSS161010 \citep{Coppejans2020}, and AT2020xnd \citep{Perley2021xnd}. 
    For comparison, the grey contours show the PTF/iPTF CCSNe host galaxy sample \citep{Schulze2021}, from $0.5\sigma$ to $2\sigma$ in steps of $0.5\sigma$. The light green and yellow bands show the main sequence of star-forming galaxies at $0.02<z<0.085$ \citep{Renzini2015} and $z\sim1$ (Eq.~4 of \citealt{Elbaz2007}), respectively. 
    \ad{\textit{Right}: The host galaxies of AT2018cow-like events and other massive star explosions on the sSFR--$M_\ast$ diagram. Histograms show the normalized distribution of 18cow-like events (thick line, unfilled), DES RETs (thin line, unfilled), ZTF FBOTs (dark filled), and SLSNe-I (light filled). }
    \label{fig:Mstar_SFR}}
\end{figure*}

The decay rate of \target is significantly slower than that of AT2018cow and AT2020xnd (Figure~\ref{fig:optlc}). 
This is similar to the post-peak decay of AT2018lug, which also slows down at $\Delta t\approx 6$--8\,days (see Figure~\ref{fig:optlc}). 
The slower decay can be caused either by the emergence of a radioactivity powered SN or continued CSM interaction. 
Since the color evolution of \target is most similar to interacting SNe shown in the upper right panel of Figure~\ref{fig:color_evol}, we slightly favor the CSM interaction scenario. 
In Appendix~\ref{sec:csm_sbo+sce}, we attempt to fit the multi-band light curve using the one-zone SBO+SCE model presented by \citet{Margalit2021}, but no satisfactory fit is obtained. However, given that the CSM interaction model has many free parameters (e.g., anisotropy, radial density structure), more detailed modeling would be required to determine whether it is a viable emission mechanism.

Assuming $T_{\rm bb} = 10^4$\,K, the bolometric luminosity and blackbody radius of \target are shown in Figure~\ref{fig:Lbol}. Although radioactivity is not required to explain the optical emission, the light curve at $\Delta t \gtrsim 10$\,days is consistent with being dominated by nickel decay with $M_{\rm ej}\sim1$--6\,$M_\odot$ and $M_{\rm Ni}\sim 0.3$--0.4\,$M_\odot$. Improved analytic relations (compared to the ``Arnett model'' shown in Figure~\ref{fig:Lbol}) have been presented by \citet{Khatami2019}. Adopting $L_{\rm peak}\approx 10^{42.8}$, $t_{\rm peak}\approx 17$\,days, and the dimensionless parameter $\beta\approx 1$, we use Eq.~21 of \citet{Khatami2019} to estimate $M_{\rm Ni}$, which gives $M_{\rm Ni}\approx 0.26\,M_\odot$. In summary, the inferred $M_{\rm ej}$ and $M_{\rm Ni}$ are broadly consistent \ad{with} stripped envelope SNe of all types (IIb, Ib, Ic, and Ic-BL; \citealt{Drout2011, Taddia2018, Prentice2019}), but can not accommodate normal hydrogen-rich type II SNe \citep{Meza2020, Afsariardchi2021}.

\subsection{A Dwarf Host Galaxy} \label{subsec:dwarf_host}

Figure~\ref{fig:Mstar_SFR} shows the position of \target on the SFR--$M_\ast$ \ad{and the sSFR--$M_\ast$ diagrams} (based on properties derived in \S\ref{subsubsec:host_analysis}). 
\ad{
For comparison, we also show the 28 FBOTs selected from ZTF (note that we excluded the three 18cow-like events from the 31 objects in Tab.~17 of \citealt{Ho2021a}), the 49 rapidly evolving transients (RETs) from the dark energy survey (DES) \citep{Wiseman2020}, and 18 PTF SLSNe-I from \citet{Perley2016slsn}.
Compared with normal CCSNe \citep{Schulze2021} and X-ray/radio-faint FBOTs, the $M_\ast$ of AT2018cow-like events (a sample of five) is much smaller. Indeed, all AT2018cow-like events} are hosted by dwarf galaxies with $M_\ast < 2\times 10^9\,M_\odot$.
This trend has been previously reported by \citet{Perley2021xnd}, and argues for a massive star origin. 
Several types of the most powerful explosions from massive stars are also preferentially hosted by dwarf galaxies, including long GRBs \citep{Vergani2015, Perley2016grb}, hydrogen-poor SLSNe \citep{Leloudas2015, Perley2016slsn, Taggart2021}, and SNe Ic-BL \citep{Schulze2021}. 

\citet{Perley2021xnd} have suggested that an elevated level of SFR or sSFR is not a requirement for producing AT2018cow and similar explosions. The properties of \target's host further support this suggestion. 
\ad{At $M_\ast \sim 10^8\,M_\odot$, the SFR of \target lies below the main-sequence (MS) of local star-forming galaxies.
Moreover, among the 369 PTF/iPTF normal CCSNe hosted by galaxies with $M_\ast < 2\times 10^9\,M_\odot$ \citep{Schulze2021}, the host galaxies of only 30 objects (8\%) have ${\rm sSFR}<8\times10^{-11}\,{\rm yr^{-1}}$. 
This indicates that \target does not occur during a vigorous starburst, and that progenitor scenarios with a slightly longer delay time than that of a typical CCSN are favored. 
\citet{Zapartas2017} performed a population synthesis study of CCSNe, finding that a prolonged delay time can be achieved by binary interactions, through common envelope evolution, mass transfer episodes, and/or merging. 
Explosions driven by the merging of a compact object with a massive star inside a common envelope have indeed been proposed as promising channels for producing AT2018cow-like events \citep{Soker2019, Schroder2020, Soker2022, Metzger2022}.
}

\ad{
Among the five AT2018cow-like events, only AT2018lug lies above the local MS of star-forming galaxies. For comparison, the majority (15/18) of SLSNe-I presented by \citet{Perley2016slsn} lie above the local MS\footnote{\ad{Compared with AT2018cow-like events, the sample of SLSNe-I is at slightly higher redshifts (the median is $z\sim 0.2$). We note that for $M_\ast \approx 10^8\,M_\odot$, the sSFR at $z\approx0.2$ is only slightly ($\approx 0.2$\,dex) higher than that at $z\approx0$ \citep{Speagle2014}.}}. 
We perform a two-sided Kolmogorov-Smirnov (K-S) test for the null hypothesis that the host galaxy sSFR of SLSNe-I and AT2018cow-like events are drawn from the same distribution. The returned $p$-value of 0.23 is too high to reject the null hypothesis. 
}
\ad{
A larger sample size is clearly needed to test if the host sSFR between AT2018cow-like events and other powerful massive star explosions are statistically different. 
}

\subsection{An Engine Driven Explosion} \label{subsec:xray_origin}

\subsubsection{X-ray Properties}

We have shown that the radio (\S\ref{subsec:shock_CSM}) and early-time optical (\S\ref{subsec:opt_spec}, \S\ref{subsec:optlc_model}) properties of \target are similar to other AT2018cow-like events. Here we summarize the key X-ray observables of \target, and compare them with other AT2018cow-like events.

At $\sim 36$\,days, the mean 0.3--10\,keV luminosity of \target is $(1.9\pm0.4)\times 10^{43}\,{\rm erg\,s^{-1}}$, a factor of $\sim 20$ brighter than AT2018cow and AT2020xnd at similar phases (Figure~\ref{fig:xlc}). The best-fit powerlaw of $f_\nu \propto \nu^{-0.8}$ (Figure~\ref{fig:srg_spec}) is similar to the 0.3--10\,keV spectral shape of AT2018cow and AT2020xnd \citep{Margutti2019, Bright2022, Ho2021b}. From 34.5 to 37.6\,days, the 0.2--2.2\,keV flux varies by a factor of $\approx 6$ on the timescale of $\approx 1$\,day (Figure~\ref{fig:srg_xlc}), similar to the fast soft X-ray variability observed in AT2018cow at similar phases (Figure~\ref{fig:xlc}).

At 328\,days, the mean 0.3--10\,keV luminosity of \target is \ad{$\sim 1.4\times 10^{42}\,{\rm erg\,s^{-1}}$}, which is $\sim300$ times brighter than the upper limit of CSS161010 at 291\,days, and $\sim200$ times brighter than AT2018cow itself at 212\,days. 
The spectrum of \target has probably hardened to $f_\nu \propto \nu^{0}$. From 327.4 to 328.2\,days, the X-ray flux decreases by a factor of \ad{$\sim 2.6$}. 

\ad{Among AT2018cow-like events, intraday X-ray variability has only been detected in AT2018cow and AT20202mrf. This is probably because CSS161010, AT2018lug, and AT2020xnd were not observed often enough to detect it. The isotropic equivalent observed X-ray luminosity of \target is as luminous as long GRBs. The X-ray emission of long GRBs are produced by the afterglow synchrotron radiation of electrons accelerated by a ultra-relativistic shock \citep{Sari1998}. However, given the lack of a prompt $\gamma$-ray emission (\S\ref{subsec:gamma-ray}) and the sub-relativistic shock velocity (\S\ref{subsec:shock_CSM}) observed in AT2020mrf, the nature of its X-rays should be different from that of long GRBs. 
}

As shown in Figure~\ref{fig:xlc}, in AT2018cow (and perhaps AT2020xnd), the 0.3--10\,keV light curve decay steepens from $L \propto t^{-1}$ ($t\lesssim25$\,days) to $L \propto t^{-4}$ ($25\lesssim t\lesssim 100$\,days). 
The overall decay shape of \target is consistent with a $L\propto t^{-1.3}$ power-law. 
However, we can not rule out the existence of a steeper decay (see \S\ref{subsubsec:BHengine}). 
Below we discuss the physical origin of the X-ray emission associated with \target. 

\begin{figure}[htbp!]
    \centering
    \includegraphics[width = \columnwidth]{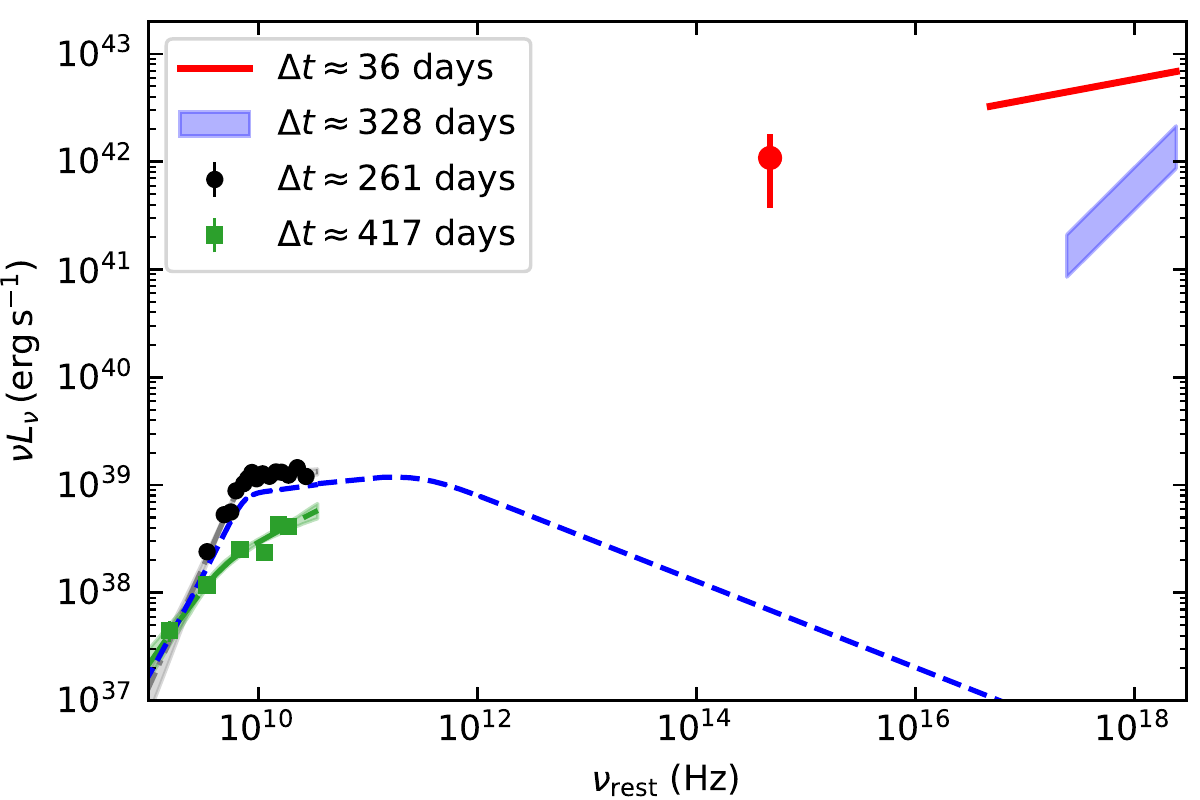}
    \caption{SED of \target. 
    \ad{The dashed blue line shows an example synchrotron spectrum one would expect at the epoch of the \chandra observation ($\Delta t\approx 328$\,days). Here we have assumed $B\sim0.2$\,G, and a cooling frequency of $\nu_c = \gamma_c^2 \nu_g\sim 3\times 10^{11}$\,Hz, where $\gamma_c = 6\pi m_e c / (\sigma_{\rm T} B^2 t)$ and $\nu_g = eB / (2\pi m_e c)$ \citep{Sari1998, Granot2002}.}
    Therefore, the late-time X-ray emission \ad{is much brighter than the synchrotron spectrum.}
    \label{fig:multiband_sed}}
\end{figure}

\subsubsection{General Considerations}

First, Figure~\ref{fig:multiband_sed} shows that the late-time X-ray luminosity of \target is too bright to be an extension of the radio synchrotron spectrum.  

Second, inverse-Compton (IC) scattering of the radiation field (i.e., UV/optical photons) by electrons accelerated in the forward shock is found to be the main early-time ($t\lesssim 40$\,days) X-ray emission mechanism for SNe Ib/c exploding in low-density environments \citep{Fransson1996, Kamble2016}.
The ratio of IC to synchrotron radiation losses is $P_{\rm IC}/P_{\rm syn} = u_{\rm rad}/ u_B$, where $u_{\rm rad}$ is the energy density in seed photons, and $u_{B} = U_B / (4\pi R_{\rm sh}^3 / 3)$. To first order, $P_{\rm IC}/P_{\rm syn}\sim L_{\rm X} / L_{\rm radio}$.
At $\Delta t \approx 36$\,days, the bolometric luminosity of the optical transient is $L_{\rm bol}\sim  10^{42.3}\,{\rm erg\,s^{-1}}$ (see Figure~\ref{fig:Lbol}). Assuming $v_{\rm sh}\sim 0.07$--0.08$c$ (\S\ref{subsubsec:C98SSA}), the shock radius is $R_{\rm sh}\sim 7\times 10^{15}$\,cm. Therefore $u_{\rm rad}/ L_{\rm X} = L_{\rm bol} / (4\pi c R_{\rm sh}^2) / L_{\rm X}\sim (0.1\,{\rm erg\,cm^{-3}}) / (2\times 10^{43}\,{\rm erg\,s^{-1}})\sim 5.4\times 10^{-45}\,{\rm s\,cm^{-3}}$. Assuming that the standard SSA model applies at $\Delta t \approx 36$\,days\footnote{This assumption will not be accurate if $v_{\rm sh}\gtrsim 0.2c$, at which condition we expect thermal electrons to contribute significantly to the synchrotron spectrum \citep{Ho2021b, Margalit2021_sync}.}, from Equation~(\ref{eq:U}) we have $u_B / L_{\rm radio} \sim 8 \times 10^{-31} L_{\theta \nu , 29}^{4/19} (\epsilon_e / \epsilon_B)^{-11/19} \nu_{100}^{-1} > 2 \times 10^{-31}\,{\rm s\,cm^{-3}}$, where we have assumed that the early-time synchrotron emission peaks at $\sim100$\,GHz and $L_{\theta \nu }>10^{29}\,{\rm erg\,s^{-1}\,Hz^{-1}}$. Therefore, $u_{\rm rad} / L_{\rm X} \ll u_{B} / L_{\rm radio}$, and IC is not likely to be the dominant mechanism for the X-ray emission.
At $\Delta t \approx 328$\,days, the observed X-ray spectral shape of $f_\nu \propto \nu^{0}$ is too hard to be consistent with IC. 

Finally, X-rays from most normal CCSNe and interacting SNe have been successfully modeled by thermal bremsstrahlung from supernova reverse-shock-heated ejecta or the forward-shock-heated CSM \citep{Chevalier1994, Dwarkadas2012}.
The shortest variability timescale expected from clumpy CSM encountered by a forward shock is much slower --- $\Delta t/ t = v_{\rm sh}/c \sim 0.1$ (see Section 3.3.1 of \citealt{Margutti2019}). In contrast, the X-ray relative variability and flux contrast are $\Delta t / t  \approx 0.03$, $\Delta F / F \approx 2.5$ at $t\approx 36$\,days and $\Delta t / t \lesssim 0.003$, $\Delta F / F \approx 1$ at $t\approx 328$\,days. 

Some previous studies have interpreted AT2018cow as the tidal disruption of a white dwarf or star by an IMBH \citep{Kuin2019, Perley2019}. Since the observed early-time non-thermal X-ray spectrum and fast variability are not consistent with observations of thermal X-ray loud TDEs \citep{Sazonov2021}, the X-rays are thought to be powered by a jet similar to that observed in the jetted TDE SwiftJ1644 \citep{Burrows2011, Bloom2011}. However, for \target and AT2018cow-like events in general, the TDE scenario is disfavored by the dense environment (\S\ref{subsec:shock_CSM}) and the host properties (\S\ref{subsec:dwarf_host}). 

Therefore, the most natural origin of the X-rays in \target is a central compact object --- either a neutron star (\S\ref{subsubsec:NSengine}) or a black hole (\S\ref{subsubsec:BHengine}) --- formed in a massive star explosion. Since the UV/optical luminosity of \target remains much lower than $L_{\rm X}$ throughout the evolution, we can assume that the central engine luminosity $L_{\rm e}$ is mostly tracked by $L_{\rm X}$. The engine timescale is set by the duration of the X-ray emission $t_{\rm e}>328$\,days. The total energy release in the X-ray is $E_{\rm e} > (2\times 10^{43}\,{\rm erg\,s^{-1}})\times (36\,{\rm days}) + (10^{42}\,{\rm erg\,s^{-1}})\times[(328-36)\,{\rm days}] = 9\times 10^{49}\,{\rm erg}$. 

\subsubsection{Stellar Mass Black Hole Engine} \label{subsubsec:BHengine} 

The engine of \target can be a stellar mass BH, where X-rays are powered by accretion. The isotropic equivalent luminosity of $10^{42}$--$10^{43}\,{\rm erg\,s^{-1}}$ corresponds to an Eddington ratio of $L_{\rm engine}/L_{\rm Edd}>10^4$--$10^3$ for a $10\,M_\odot$ BH, suggesting that the emission is likely beamed.

In the case of a failed explosion, $t_{\rm e}$ is determined by the free-fall of the stellar envelope \citep{Quataert2012, Fernandez2018}:
\begin{align}
    t_{\rm ff} = \frac{\pi r^{3/2}}{(2GM_\star)^{1/2}} = 706 \left(\frac{r}{10^{14}\,{\rm cm}} \right)^{3/2} \left( \frac{M_\star}{10\,M_\odot}\right)^{-1/2}\,{\rm day}.
\end{align} 
In order to power \target's X-ray emission out to 328\,days, a weakly bound red supergiant (RSG) progenitor with $r>6\times10^{13}\,{\rm cm}$ is required. The amount of mass around the disk circularization radius is much smaller than that in the stellar envelope, and the fast X-ray variability is related to the change of angular momentum in the accreting material \citep{Quataert2019}.

In the case of a successful explosion, the accretion is supplied by fallback of bound material \citep{Dexter2013}. In compact progenitors such as blue supergiants (BSGs), a reverse shock decelerates the inner layers of the ejecta, resulting in enhanced fallback mass \citep{Zhang2008}. 
The fast X-ray variability might be caused by disk instability since the viscous time
is much shorter than the fallback time. 
The temporal coverage of our X-ray data is poor. It is possible that $L_{\rm e}$ decays shallower than $t^{-1.3}$ initially, followed by a steeper decay (e.g. $L_{\rm e} \propto t^{-5/3}$) due to fallback. This might be consistent with a range of SN energies, with lower energies corresponding to later transition times between an early less steep light curve to a later steeper fallback light curve \citep{Quataert2012}. 

\subsubsection{Millisecond Magnetar Engine} \label{subsubsec:NSengine}

Another speculation is that the engine of \target is a young magnetar (i.e., an extremely magnetized neutron star), where $L_{\rm e}$ is primarily provided by rotational energy loss due to spindown. For a neutron star with a spin period of $P_{\rm ms}\equiv P / (1\,{\rm ms})$ and a mass of $1.4\,M_\odot$, the rotational energy is $E_{\rm rot} \approx 2.5\times 10^{52} P_{\rm ms}^{-2}\,{\rm erg}$ \citep{Kasen2010, Kasen2017}. The spin period required to power $E_{\rm e}$ is thus $P\lesssim 17$\,ms. If the NS has a radius of $10$\,km and a magnetic field of $B_{14} \equiv B / (10^{14}\,{\rm G})$, the characteristic spindown timescale is $t_{\rm spindown}\approx 0.5 B_{14}^{-2} P_{\rm ms}^2$\,day. The luminosity extracted from spindown is roughly constant when $t \lesssim t_{\rm spindown}$, and decays as $L_{\rm e}\propto t^{-2}$ afterwards. Extrapolating the \chandra detection back to the \srg luminosity suggests that the transition occurs at $\sim 73$\,days, which implies $B\lesssim 1.4\times 10^{14}$\,G. This is similar to the $B$ field required to power AT2018cow inferred by \citet{Margutti2019}. 

In this scenario, X-rays are generated in a ``nebula'' region of electron/positron pairs and radiation inflated by a relativistic wind behind the SN ejecta \citep{Vurm2021}. Additional energy injection by fallback accretion widens the parameter space of magnetar birth properties, and predicts a late-time light curve decay shallower or steeper than $L_{\rm e}\propto t^{-2}$ \citep{Metzger2018}. The day-timescale X-ray variability can be accounted for by magnetically driven mini-outbursts.

\section{The Detection Rate in X-ray Surveys} \label{sec:rate}
\begin{deluxetable*}{c|c|c|c|c|c}[htbp!]
	\tablecaption{The detection rates ($\dot N_{\rm det}$ in yr$^{-1}$) of events similar to AT2018cow and AT2020mrf in X-ray surveys, under three different assumptions of the event volumetric rates ($\mathcal{R}$ in ${\rm Gpc^{-3} \,yr^{-1}}$).\label{tab:rate}}
	\tablehead{
		\colhead{Survey}
		& \colhead{$f_{-13}$}
		& \colhead{$D_{\rm max}$}
		& \colhead{$\dot N_{\rm det}$ if $\mathcal{R}=2.1$}
		& \colhead{$\dot N_{\rm det}$ if $\mathcal{R}=70$} 
		& \colhead{$\dot N_{\rm det}$ if $\mathcal{R}=420$} 
	}
	\startdata
	\hline
	\multirow{2}{*}{\srg/eROSITA}  & \multirow{2}{*}{1.8} 
	& \cellcolor{yellow}373 
	& \cellcolor{yellow}0.080 
	& \cellcolor{yellow}2.7 
	& \cellcolor{yellow}16 \\
	 &  & \textbf{964} & \textbf{1.7} & \textbf{57} & \textbf{340} \\
	\hline
	\multirow{2}{*}{Einstein Probe} & \multirow{2}{*}{20} 
	& \cellcolor{yellow}112 
	& \cellcolor{yellow}0.012 
	& \cellcolor{yellow}0.41 
	& \cellcolor{yellow}2.5\\
	& & \textbf{289} & \textbf{0.21} & \textbf{7.1} & \textbf{43}\\
    \enddata
    \tablecomments{$D_{\rm max}$ is given in Mpc. The values with yellow background assume an X-ray light curve shape similar to AT2018cow. The values given in boldface assume a conservative light curve shape similar to AT2020mrf, and therefore the derived $\dot N_{\rm det}$ should be taken as lower limits.}
\end{deluxetable*}

\target is the first multi-wavelength FBOT identified from X-ray surveys. This motivates us to estimate the rate of such events in present and future X-ray surveys. The core collapse SN rate is $R = 7 \times 10^4 \,{\rm Gpc^{-3} \,yr^{-1}}$ \citep{Li2011a}. The birthrate of 18cow-like events estimated by ZTF is $\mathcal{R} = 3\times 10^{-5}$--$6\times 10^{-3}R$ \citep{Ho2021a}, or 2.1--$420\,{\rm Gpc^{-3} \,yr^{-1}}$.

Here we assume that a multi-wavelength FBOT has an X-ray light curve either similar to AT2018cow itself or similar to AT2020mrf. 
We approximate the 0.3--10\,keV X-ray luminosity of AT2018cow as a plateau with a luminosity of $L_{\rm X, p0}= 3 \times 10^{42}\,{\rm  erg \,s^{-1}}$ and a duration of $t_{\rm X, p0} =30$\,days (Figure~\ref{fig:xlc}).
The light curve shape of AT2020mrf is less well constrained. 
For simplicity, we assume a conservative shape consisting of two plateaus, with $L_{\rm X, p1} = 2\times 10^{43}\,{\rm erg\,s^{-1}}$, $t_{\rm X, p1} =36$\,days, $L_{\rm X, p2} = 1\times 10^{42}\,{\rm erg\,s^{-1}}$, and $t_{\rm X, p2} =350$\,days.

The transient detection rate is 
\begin{align}
    \dot N_{\rm det} = \frac{\Omega}{3} D_{\rm max}^3 \mathcal{R} \cdot p_{\rm s}
\end{align}
where $\Omega$ is the solid angle of the surveyed area ($\Omega = 4\pi$ for an all-sky survey), $D_{\rm max}$ is the maximum distance out to which the source can be detected, and $p_{\rm s}$ is the probability that the transient is ``on'' when being scanned by the X-ray survey. 
If the survey cadence is shorter than the transient duration, $p_{\rm s} = 1$.
Setting a survey flux threshold of $f_{\rm thre} = 10^{-13} f_{-13}\,{\rm  erg \, cm^{-2}\, s^{-1}}$, we have $4\pi D_{\rm max}^2 f_{\rm thre} = L_{\rm X, p}$. 

On average, every 0.5\,yr, \srg/eROSITA samples the same region of the sky over $\sim12$\,passes within $\sim2$\,days. 
For a single event, somewhere on the sky, with an X-ray light curve shape similar to AT2018cow, the probability of being imaged by \srg during its X-ray active phase is $p_{\rm s0} = 2\times(t_{\rm X, p0}+2)/365 = 0.175$.
For a light curve shape similar to AT2020mrf, $p_{\rm s1} = 2\times(t_{\rm X, p1}+2)/365 = 0.208$, and $p_{\rm s2} \sim 1$.
The sensitivity of an eROSITA single sky survey is $\approx 2.5\times 10^{-14}\,{\rm erg\,s^{-1}\,cm^{-2}}$ (see Fig.~17 of \citealt{Sunyaev2021}). In reality, to be selected as a transient by eRASS$n$ ($n>1$), the source needs to exceed the eRASS1 sensitivity limit by a factor of $\approx 7$. Therefore, the flux threshold is $f_{-13}\approx 1.8$. 

Einstein Probe (EP) is a lobster-eye telescope for monitoring the X-ray sky \citep{Yuan2018} to be launched at the end of 2022. 
With an orbital period of 97\,min, the entire sky can be covered over three successive orbits.  
Here we assume that its Wide-field X-ray telescope (WXT) is 2 orders of magnitude more sensitive than the Monitor of All-sky X-ray Image (\maxi) mission\footnote{From slide \#32 of \url{https://sites.astro.caltech.edu/~srk/XC/Notes/EP_20200923.pdf}}. 
\maxi has a transient triggering threshold of 8\,mCrab for 4\,days \citep{Negoro2016}, leading us to assume $f_{-13}\approx 20$ for EP.

The calculated detection rates in eROSITA and EP are summarized in Table~\ref{tab:rate}. The rate of similar events in present and future millimeter transient surveys is given by \citet{Ho2021b} \ad{and \cite{Eftekhari2021}}.


\section{Summary} \label{sec:conclusion}

We report multi-wavelength observations of \target, the fifth member of the class of AT2018cow-like events (i.e., FBOTs with luminous multi-wavelength counterparts). 
Among the four 18cow-like events ever detected in the X-ray (i.e., AT2018cow, CSS161010, AT2020xnd, AT2020mrf), \target is the most luminous object, exhibiting day-timescale X-ray variability both at early ($\approx 36$\,days) and late times ($\approx 328$\,days), with a luminosity between $10^{42}$ and ${\rm few}\times 10^{43}\,{\rm erg\,s^{-1}}$. 
Previously, the only object showing evidence of a NS/BH central engine was AT2018cow \citep{Margutti2019, Pasham2021}. Here we show that a compact object --- a young millisecond magnetar or an accreting black hole --- is required to be the central energy source of AT2020mrf (see \S\ref{subsec:xray_origin}).

\target also provides accumulating evidence to show that AT2018cow-like events form another class of engine-driven massive star explosions, after long GRBs and SLSNe-I. Intriguingly, all three classes of events are preferentially hosted by dwarf galaxies. Given the MZR \citep{Gallazzi2005, Berg2012, Kirby2013}, low metallicity probably plays an important role in the formation of such exotic explosions by reducing angular momentum loss 
of their progenitors \citep{Kudritzki2000}. 
Local environment studies with integral-field unit (IFU) observations (e.g., \citealt{Lyman2020}) and high spatial resolution images (e.g., with the Hubble Space Telescope) can further illuminate the nature of their progenitors.

Although AT2018cow, AT2018lug, and AT2020xnd are FBOTs with $-20.5< M_{g, \rm peak}< -21.5$ and $t_{1/2}< 5$\,day, the optical light curve of \ad{AT2020mrf} is of lower peak luminosity ($M_{g, \rm peak}=-20$) and slower evolution timescale ($t_{1/2}=7$\,days). This should guide searches of such events in optical wide field surveys to be more agnostic of the light curve decay rate. Real-time identification of FBOTs and comprehensive spectroscopic follow up observations are necessary to distinguish between different emission mechanisms: shock interaction with extended CSM, radioactivity, or wind reprocessing (\S\ref{subsubsec:opt_power}). The discovery of X-ray emission in \target also showcases how X-ray surveys such as \srg can be essential in the identification of multi-wavelength FBOTs. 

Once identified, millimeter and radio follow-up observations are needed to reveal the CSM density as a function of distance to the progenitor, which contains information about the mass-loss history (\S\ref{subsec:shock_CSM}). X-ray light curves provide diagnostics for the nature of the power source (\S\ref{subsec:xray_origin}), while broad-band X-ray spectroscopy can constrain the evolution of the geometry of the material closest to the central engine \citep{Margutti2019}. Given the late-time X-ray detections of AT2018cow at $\Delta t \approx 212$\,days (Appendix~\ref{sec:18cow_xmm}) and of AT2020mrf at $\Delta t \approx 328$\,days (\S\ref{subsec:cxo}), future \chandra observations of these two objects may further constrain the timescales of their central engines.


\vspace{1cm}

\textit{Acknowledgements} -- 
We thank Patrick Slane for allocating DD time on \chandra. We thank the staff of \chandra, VLA, Keck, and GMRT that made these observations possible. 
We thank Jim Fuller, Mansi Kasliwal, Wenbin Lu, Tony Piro, and Eliot Quataert for helpful discussions. \ad{
We thank the anonymous referee for constructive comments and suggestions.
Y.Y. thanks Eric Burns for discussion about IPN, and Dmitry Svinkin for providing information about Konus-WIND.}

\ad{Support for this work was provided by the National Aeronautics and Space Administration (NASA) through Chandra Award Number DD1-22133X issued by the Chandra X-ray Observatory Center, which is operated by the Smithsonian Astrophysical Observatory for and on behalf of NASA under contract NAS8-03060.}

Y.Y. acknowledges support by the Heising-Simons Foundation. 
Nayana A.J. would like to acknowledge DST-INSPIRE Faculty Fellowship (IFA20-PH-259) for supporting this research. 
P.C. acknowledges support of the Department of Atomic Energy, Government of India, under the project no. 12-R\&D-TFR-5.02-0700. 
P.M., M.G., S.S., G.K. and R.S. acknowledge the partial support of this research by grant 21-12-00343 from the Russian Science Foundation.

GMRT is run by the National Centre for Radio Astrophysics of the Tata Institute of Fundamental Research.

This work is based on observations with the eROSITA telescope on board the \srg observatory. The \srg observatory was built by Roskosmos in the interests of the Russian Academy of Sciences represented by its Space Research Institute (IKI) in the framework of the Russian Federal Space Program, with the participation of the Deutsches Zentrum f\"{u}r Luft- und Raumfahrt (DLR). The \srg/eROSITA X-ray telescope was built by a consortium of German Institutes led by MPE, and supported by DLR. The \srg spacecraft was designed, built, launched and is operated by the Lavochkin Association and its subcontractors. The science data are downlinked via the Deep Space Network Antennae in Bear Lakes, Ussurijsk, and Baykonur, funded by Roskosmos. The eROSITA data used in this work were processed using the eSASS software system developed by the German eROSITA consortium and proprietary data reduction and analysis software developed by the Russian eROSITA Consortium. 

The ZTF forced-photometry service was funded under the Heising-Simons Foundation grant \#12540303 (PI: Graham). This work has made use of data from the ATLAS project, which is primarily funded to search for near earth asteroids through NASA grants NN12AR55G, 80NSSC18K0284, and 80NSSC18K1575; byproducts of the NEO search include images and catalogs from the survey area. This work was partially funded by Kepler/K2 grant J1944/80NSSC19K0112 and HST GO-15889, and STFC grants ST/T000198/1 and ST/S006109/1. The ATLAS science products have been made possible through the contributions of the University of Hawaii Institute for Astronomy, the Queen’s University Belfast, the Space Telescope Science Institute, the South African Astronomical Observatory, and The Millennium Institute of Astrophysics (MAS), Chile.


\software{
\texttt{astropy} \citep{Astropy2013},
\texttt{CASA} (v5.6.1; \citealt{McMullin2007}), 
\texttt{CIAO} \citep{Fruscione2006},
\texttt{emcee} \citep{Foreman-Mackey2013},
\texttt{LPipe} \citep{Perley2019lpipe}, 
\texttt{matplotlib} \citep{Hunter2007},
\texttt{pandas} \citep{McKinney2010},
\texttt{Prospector} \citep{Johnson2021},
\texttt{pyne2001} \citep{Cordes2002},
\texttt{python-fsps} \citep{Foreman-Mackey2014},
\texttt{scipy} \citep{Virtanen2020}
}

\facilities{
CXO, XMM, 
PO:1.2m, 
Keck:I (LRIS), 
VLA, GMRT
}

\appendix

\section{\xmm Late-time Detection of AT2018cow} \label{sec:18cow_xmm}

\begin{figure}[htbp!]
    \centering
    \includegraphics[width = 0.5\columnwidth]{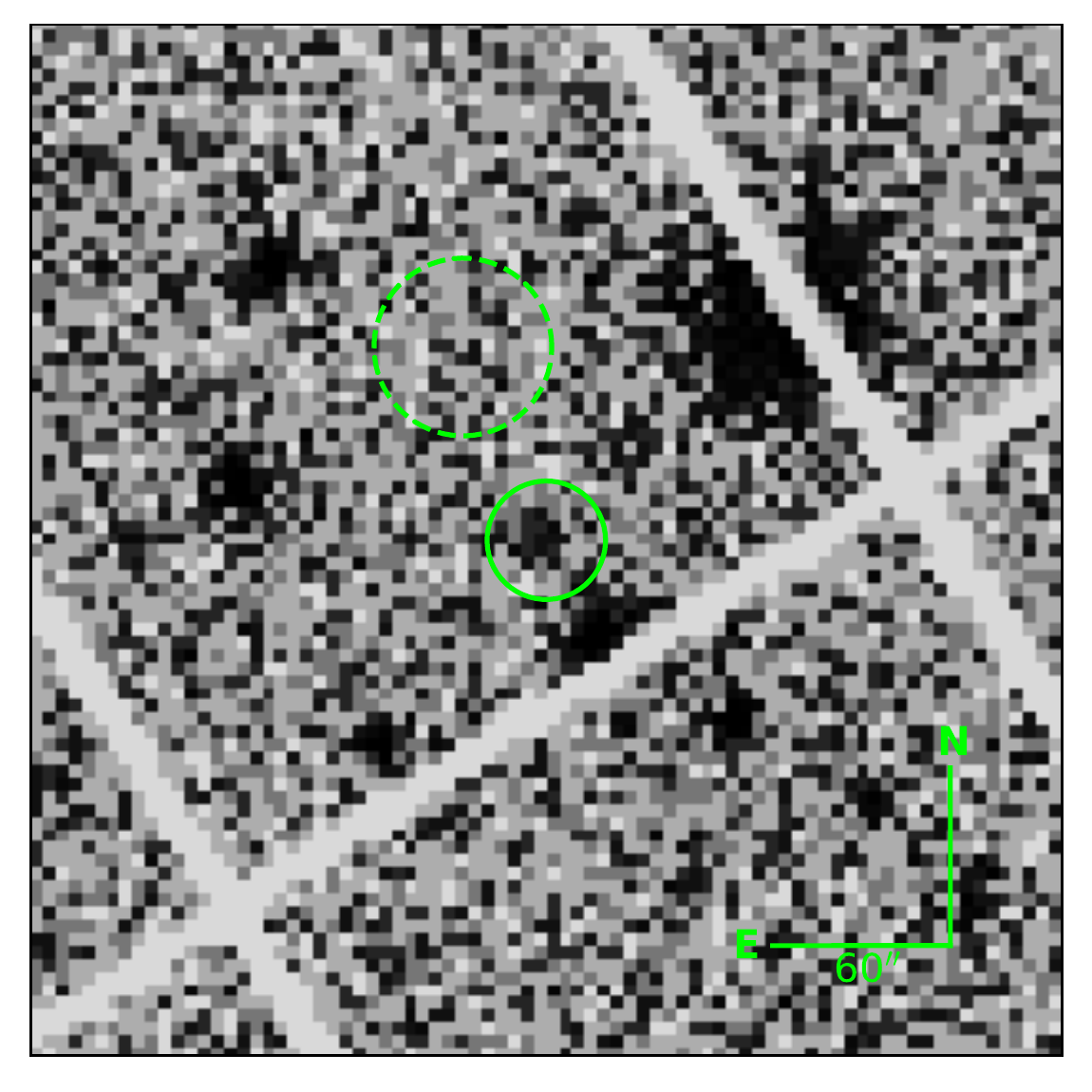}
    \caption{\xmm/pn 0.3--10\,keV image centered on AT2018cow, obtained at $\Delta t = 212$\,days. The solid circle is the source region with $r_{\rm src} = 20^{\prime\prime}$, and the dashed circle is the background region with $r_{\rm bkg} = 30^{\prime\prime}$. \label{fig:18cow_xmm3}}
\end{figure}

AT2018cow was observed by \xmm/EPIC on three epochs (PI Margutti) at rest-frame 29.6, 78.1, and 211.8\,days since explosion. The first two epochs yielded clear X-ray detections, which have been reported by \citet{Margutti2019}. \citet{Pasham2021} analyzed the 0.25--2.5\,keV EPIC/MOS1 data of the third epoch, and reported a non detection. Here we analyze the third epoch EPIC/pn data to derive the flux (or upper limit) in 0.3--10\,keV, which is important to be compared with the late-time X-ray detection of AT2020mrf. The pn instrument generally has better sensitivity than MOS1 and MOS2.

We reduced the pn data using the \xmm Science Analysis System (SAS) and relevant calibration files.
Events were filtered with the conditions \texttt{PATTERN<=4} and \texttt{(FLAG\&0xfb0825)==0}. We removed high background time windows and retained 43178\,s good times among the total exposure time of of 53163\,s. Following \citet{Margutti2019}, we extracted the source using a circular region with a radius of $r_{\rm src} = 20^{\prime\prime}$ to avoid contamination from a nearby source located $36.8^{\prime\prime}$ southwest form AT2018cow. The background is extracted from a source-free circular region with a radius of $r_{\rm bkg}=30^{\prime\prime}$ on the same CCD (see Figure~\ref{fig:18cow_xmm3}). 

\begin{figure}[htbp!]
    \centering
    \includegraphics[width = 0.7\columnwidth]{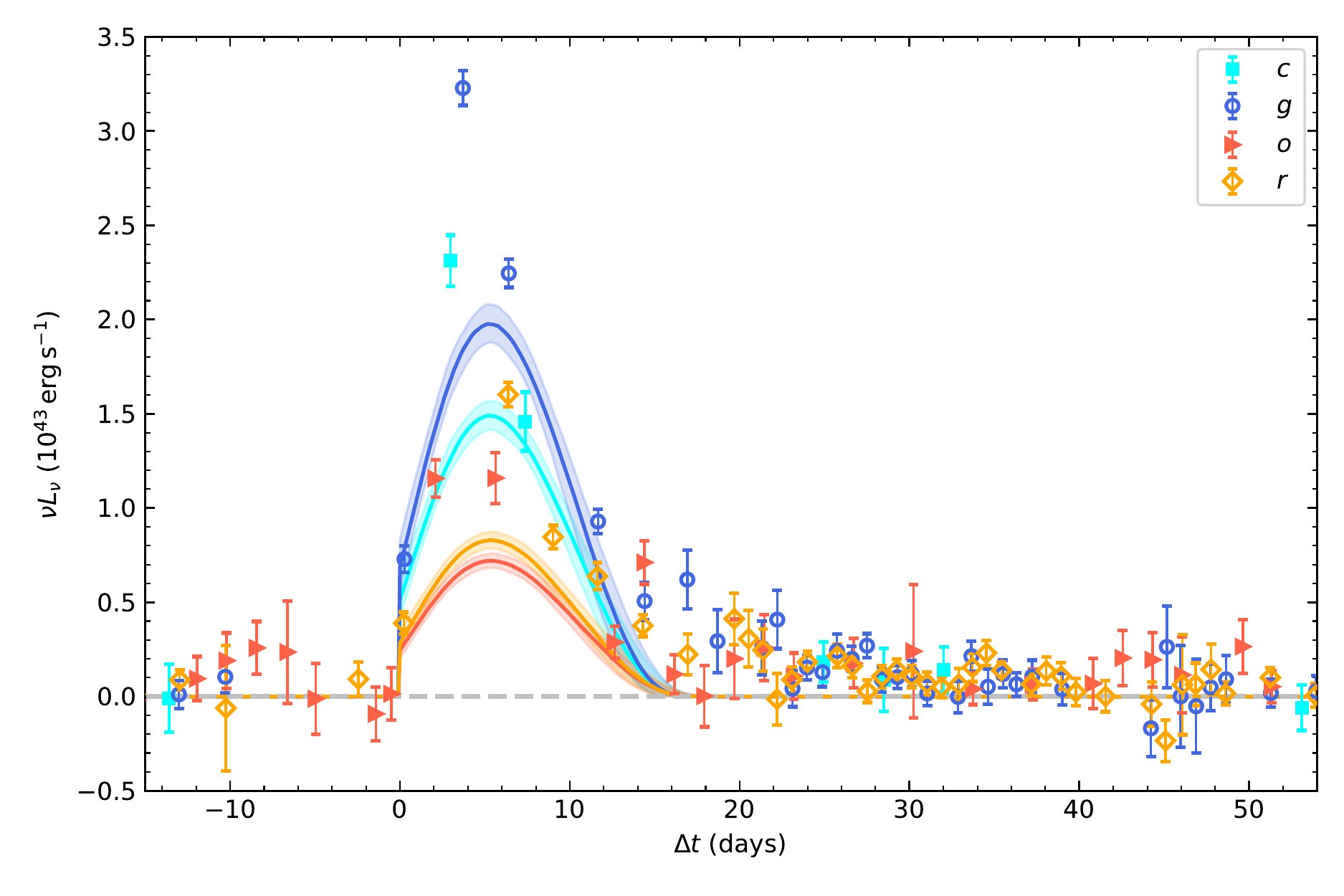}
    \caption{Dense CSM shock breakout and cooling model fit to the multi-band light curve of \target. The maximum a posteriori model is shown via solid lines. \label{fig:csm_m19_bestfit}}
\end{figure}

The average count rate of the source is $0.00486\,{\rm count\,s^{-1}}$. The average count rate of the background (multiplied by $r_{\rm src}^2/r_{\rm bkg}^2$ to match the area of the source region) is $0.00360\,{\rm count\,s^{-1}}$. Therefore, AT2018cow is detected at a (Gaussian equivalent) confidence limit of $4.2\sigma$. Assuming an absorbed power-law model with $\Gamma\approx 2$ and $N_{\rm H} \approx 7\times 10^{20}\,{\rm cm^{-2}}$, the 0.3--10\,keV flux is $\sim 1.6\times 10^{-14}\,{\rm erg\,s^{-1}\,cm^{-2}}$, corresponding to a luminosity of $\sim 7\times 10^{39}\,{\rm erg\,s^{-1}}$.

\section{A Sample of GRB X-ray Light Curves} \label{sec:grb_xlc}
The sample of GRB light curves shown in Figure~\ref{fig:xlc} is collected as follows. We start with the list of GRBs given by the \swift GRB Table\footnote{\url{https://swift.gsfc.nasa.gov/archive/grb_table/fullview/}.}. Next, we retain the 339 long GRBs ($T_{90}>2\,{\rm s}$) with reported redshifts. After that, we require the last \swift/XRT detection to be at \ad{$[(t-T_0)/(1+z)] >20$\,days}, where $T_0$ is the GRB trigger time. This step selects 12 events, including   
  \ad{GRB171205A ($z=0.0368$)}, 
  GRB190829A ($z = 0.078$), 
  GRB180728A ($z = 0.12$), 
  GRB161219B ($z = 0.15$), 
  GRB130427A ($z = 0.34$), 
  GRB061021 ($z = 0.35$), 
  GRB091127 ($z = 0.49$), 
  GRB060729 ($z = 0.54$), 
  \ad{GRB090618 ($z = 0.54$)}, 
  GRB090424 ($z = 0.54$), 
  GRB080411 ($z = 1.0$), 
  and GRB100814A ($z = 1.4$). 
We supplement the XRT light curves with deep late-time X-ray observations reported in the literature \citep{Grupe2010, DePasquale2017}.
 
\section{Modeling the Optical Light Curve with CSM SBO+SCE} \label{sec:csm_sbo+sce}

\begin{figure}[htbp!]
    \centering
    \includegraphics[width = 0.7\columnwidth]{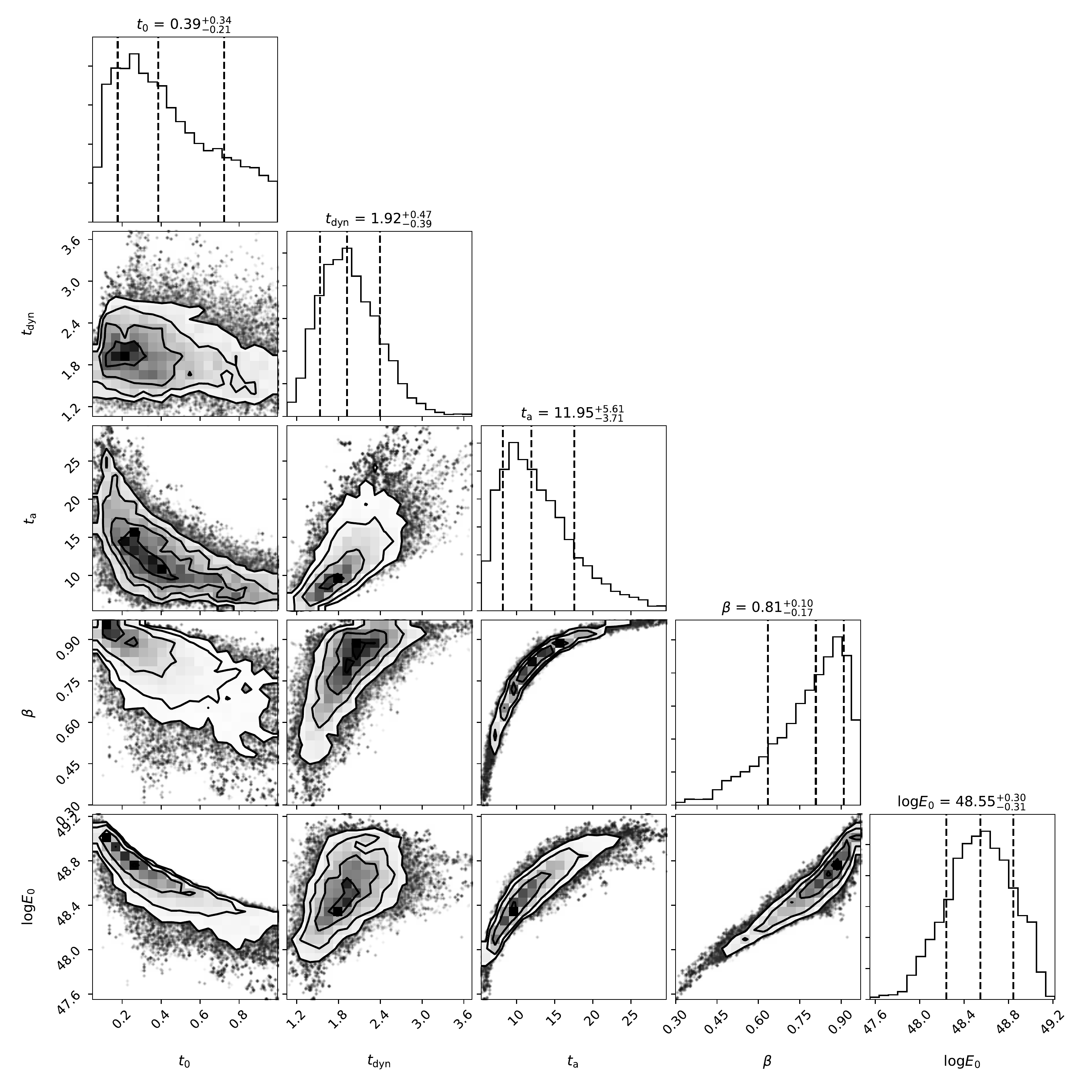}
    \caption{Corner plot showing the posterior constraints on the model parameters. Each parameter is marginalized over $\sigma_0$. \label{fig:corner_csm}}
\end{figure}

For simplicity, we adopt the one-zone model presented in Appendix~A of \citet{Margalit2021} to fit the optical light curve of \target. Following \citet{Yao2019}, we add a constant additional variance $\sigma_0^2$ to each of the measurement variance $\sigma_i^2$ to account for systematic uncertainties. The multi-band light curves are parameterized using five free parameters: $t_0$, $t_{\rm dyn}$, $t_{\rm a}$, $\beta$, and $E_0$ (see Table~1 of \citealt{Margalit2021} for the definitions of these variables). The best-fit model is shown in Figure~\ref{fig:csm_m19_bestfit}, and the posterior distribution is shown in Figure~\ref{fig:corner_csm}. 

We are not able to obtain a decent fit to the observed light curves. This is due to the fact that in the CSM shock breakout and cooling model, the light curve decay can not be significantly slower than the rise, making it difficulty to reproduce the ``flux excess'' observed at $\Delta t\sim$15--35\,days.

\bibliography{main}{}
\bibliographystyle{aasjournal}

\end{document}